\documentclass[a4paper,12pt]{article}
\usepackage{epsfig}
\usepackage{graphicx}
\usepackage{dcolumn}
\usepackage{bm}
\usepackage{longtable}
\usepackage{amssymb}
\usepackage{amsfonts}
\usepackage[margin=1.0in]{geometry}
\usepackage{hyperref}
\usepackage{amsmath}
\usepackage{multirow}
\allowdisplaybreaks[1]

\def\dd{\textrm{d}}  				     %%

\newcommand{\be}{\begin{equation}}
\newcommand{\ee}{\end{equation}}
\newcommand{\one}{{\rm 1\kern -.9mm l}} %%

\usepackage{slashed}

\newcommand{\ft}[2]{{\textstyle\frac{#1}{#2}}}

\def\e{\epsilon}

\def\l{\lambda}

\def\m{\mu}

\def\bN{\,\mathbf{N}}
\def\bM{\,\mathbf{M}}
\def\bK{\,\mathbf{K}}
\def\bV{\,\mathbf{V}}
\def\bB{\mathbf{B}}

\makeatletter							  %%
\@addtoreset{equation}{section}					  %%
\makeatother							  %%
\begin{document}
\begin{titlepage}
\begin{flushright}
DFTT 8/2012\\
DISIT-2012\\
DFPD-12/TH/6

\par\end{flushright}
\vskip 1.5cm
\begin{center}
\textbf{\huge \bf Fermionic Wigs for \\ \vspace{.3cm} ${\rm AdS}$-Schwarzschild Black Holes}
\textbf{\vspace{2cm}}\\
{\Large L.G.C.~Gentile$ ^{~a, c, e,}$\footnote{lgentile@pd.infn.it},$\ $ P.A.~Grassi$ ^{~a, d,}$\footnote{pgrassi@mfn.unipmn.it}  and A.~Mezzalira$ ^{~b, d,}$\footnote{mezzalir@to.infn.it}}

%\vfill{}
\begin{center}
{a) { \it DISIT, Universit\`{a} del Piemonte Orientale,
}}\\
{{ \it via T. Michel, 11, Alessandria, 15120, Italy, }}
 \\ \vspace{.2cm}
 {b) { \it Dipartimento di Fisica Teorica, Universit\`a di Torino,}}\\
 {{\it via P. Giuria, 1, Torino, 10125, Italy}}
\\ \vspace{.2cm}
{c) { \it Dipartimento di Fisica Galileo Galilei,\\
Universit\`a di Padova,\\
Via Marzolo 8, 35131 Padova, Italy
}}
 \\ \vspace{.2cm}
 {d) { \it INFN - Gruppo Collegato di Alessandria - Sezione di Torino}}
 \\ \vspace{.2cm}
 {e) { \it INFN, Sezione di Padova,\\
Via Marzolo 8, 35131, Padova, Italy}}
\end{center}

\par\end{center}
\vfill{}

\begin{abstract}
{\vspace{.3cm}

       \noindent
We provide the metric, the gravitino fields and the gauge fields to all orders in the
fermionic zero modes for $D=5$ and $D=4, N=2$ gauged supergravity solutions
starting from non-extremal $AdS$--Schwarzschild black holes. We compute the Brown-York
stress--energy tensor on the boundary of $AdS_5 / AdS_4$ spaces and we
discuss some implications of the fermionic corrections to perfect fluid interpretation
of the boundary theory. The complete non-linear solution, which we denote as fermionic wig,
is achieved by acting with
supersymmetry transformations upon the supergravity fields and that expansion naturally
truncates at some order in the fermionic zero modes.

}
\end{abstract}
\vfill{}
\vspace{1.5cm}
\end{titlepage}

\vfill
\eject

\tableofcontents
%\newpage
\setcounter{footnote}{0}

%%%%%%%%%%%%%%%%%%%%%%%%%%%%%%%%%%%%%%%%%%%%%%%%%%%%%%%%%%%%%%%%%%%%%%%%%%%%%%%%%%%%%%%%%%%%%%%%%%%%%%%%%%%%%%%%%%%%%%%%

\section{Introduction}

Recently, we computed the corrections to Navier-Stokes (NS) equations
due to the fermionic superpartner of a non-extremal black hole in $N=2$, $D=5$ supergravity
\cite{Gentile:2011jt}. The technique is based on seminal work \cite{Bhattacharyya:2008jc,Rangamani:2009xk}.
Here we consider the following situation:
we start form a $AdS_5$-Schwarzschild black hole solution of $D=5$ supergravity which breaks all supersymmetries
preserving only seven isometries of the $AdS$ space;
then using the $AdS$-Killing spinors we perform a
 supersymmetry transformation of the metric where the gravitino field is generated by the Killing spinors. The metric acquires new terms which are proportional to the fermionic
bilinears and in terms of those we computed the modifications to the classical relativistic NS equations
for a conformal fluid on the boundary of $AdS$ space. Unfortunately, this is not enough to derive the complete
non-linear NS equations since a finite supersymmetry transformation is needed in order to
compute the full result. Again following \cite{Bhattacharyya:2008jc,Rangamani:2009xk}, one has to construct the variation of the metric
under a finite isometry (or superisometry), which satisfies the Einstein equations, then allowing
the parameters of the isometry to become dependent upon the coordinates on the boundary, one can
derive the equations of motion corresponding to NS equations. To repeat the program for supersymmetry,
we have to construct a finite transformation, but in that case due to the anticommuting
nature of supersymmetry parameters, the series truncates after few steps.
The unconventional nature of fermionic hair prompted us to adopt the word ``wigs'' to denote the Schwarzschild solution decorated with fermionic zero modes.
The number of needed steps
depends upon the number of independent fermionic parameters entering the supersymmetry
transformations, therefore in our case it depends upon the number of the independent parameters of the $AdS$ Killing spinors.

We can change the perspective and we can look at the problem in the following way: given a bosonic solution of supergravity field equations, one can compute the zero modes of the
fermionic field equations ($3/2$- and $1/2$-spin fields). Those solutions are the
components of a supermultiplet and they trasform into themselves under supersymmetry
transformations. This can be easily seen at the quadratic level, namely, by taking into
account fermionic quadratic terms of the action or, equivalently, linear fermionic field
equations. Nonetheless, those solutions can be extended at the non-linear level by
considering all terms of the lagrangian and by expanding the solution in terms
of fermionic fields. That has an incredible advantage over the a solution with
bosonic hair (see for example for a recent development along that line
\cite{Dias:2011tj}) since the fermionic wigs are automatically trimmed by their fermionic
nature.

Based on \cite{Aichelburg:1986wv,Aichelburg:1987hy,Aichelburg:1987hx,Aichelburg:1987hz,Aichelburg:1987ia}, we construct
the complete solution of the supergravity equations. We start from a Schwarzschild--type solution, breaking
all supersymmetries and preserving 7 isometries of the $AdS_5$ background. The metric depends upon the coordinate $r$
measuring the distance between the center of $AdS_5$ space and the boundary. We choose a flat $D=4$ boundary.
Notice that Lorentz symmetry is manifestly broken by our solutions since the time is treated differently
from $3d$ space coordinates. With the factorization of the metric into a $2d$ space-time $(r,t)$ and
$3d$ space $(x^i)$, we can factorize the spinors into corresponding irreducible representations.
We compute the $AdS_5$ Killing spinors and we see that there are two independent choices
which are relevant for our study. Then, we compute the variation of the gravitino fields under the supersymmetry
where the parameters are replaced by the Killing spinors. That produces the first term of the
fermionic expansion of the gravitino solutions to the Rarita-Schwinger equation of motion.
The next step is
to compute the second variation of the metric in terms of fermionic bilinears $(\lambda, \bN, \bK_i)$.
That is achieved by computing the second supersymmetry variation of the metric.
At this stage one can check whether the Einstein equations are indeed satisfied.
We compute then the effect of the interactions to the RS equations due to fermions and to the coupling of fermions
to bosons.
Already at this step, the usage of Fierz identities to rearrange the bilinears is essential to reduce all possible
terms.
The iteration proceeds until the number of independent fermions truncates the series.  In the process, the
gauge field (the graviphoton), which has been set to zero from the beginning, is generated and its field is proportional to the fermion bilinears.
We check also the Maxwell equations order-by-order.

The computation of the Killing spinors reveals that there are essentially two structures to be taken into account (in the
text we denote those contributions as $\eta_0$ and $\eta_1$). In the first case the complete solution obtained by resumming
all fermionic contributions is rather simple since the dependence upon the boundary coordinates is very mild. On the contrary
the computations of the complete metric in the case of $\eta_1$ is rather length since all possible structures are eventually generated.
In addition, the two structures, at a certain point, start to mix and therefore a long computation has to be
done. 
This is due to the fact that by breaking Lorentz invariance from the beginning all terms of the spin connection,
of the vielbeins and of the gauge fields are generated.
Therefore we cannot use covariance under Lorentz transformation to cast our computation in an elegant and compact form and, generically,
all components are different from zero.
Technically, in order to re-sum all contributions we compute the full solution using \verb Mathematica \textsuperscript{\textregistered}.
The result is provided in a form which is still difficult to read (the electronic notebook with the D=4 and D=5 solutions is provided
as ancillary files of the preprint publication).
Nevertheless, we make some remarks regarding the results and we give the explicit formulas for the simplest cases.

Our construction has different purposes. First of all, we will use the present results
for deriving the complete non-linear Navier-Stokes equations with fermionic contributions \cite{Gentile:2011jt}.
That would be the natural final aim of the present work, but since the results  are independent from that, we decided
to present the derivation of NS equations in a separate paper. Second, the natural question is whether
the same analysis can be done also in the case of BPS solutions. For that we refer to the first
step given in \cite{Behrndt:1998jd} and we will complete their constructions by our
 algorithm. 
Another question is the case of $D=4$. In that
case a complete explicit solution is attainable and we will publish this result elsewhere.
Finally, an issue that can be addressed with our computation is the presence of
ghost modes in construction of \cite{Hristov:2010eu}.

In sec. 2, we summarize the main ingredients of $D=5$ and $N=2$ supergravity and we list the
choices we made to build our complete solution. Notice that the solution we are considering is
suitable also for $D=4$ and $N=2$ and therefore we provide the complete solution also in that case.
We also provide some comments about spinor relations and supersymmetry transformations.
In sec. 3 we discuss the Killing vectors of $AdS_5$ in our coordinate system and the boosted solution and some
considerations regarding  the choice of the coordinate system. In sec. 4, according to the precedent section we
compute the Killing spinors. Finally, in sec. 5 we discuss the algorithm and in sec. 6 we compute the metric wigs.
There, we show that even though the metric explicitly depends upon the fermion bilinears, some macroscopic quantities
such as the ADM mass do not. The complete computation is obtained in the case for $\eta_1$.
In sec. 7, we compute the boundary stress-energy tensor as the starting point for the NS equations.
 In appendices we collect some auxiliary material.

\section{Truncated $N=2$, $D=5$ Gauged Supergravity}

We provide some useful ingredients  for our computation based on papers \cite{D'Auria:1981kq,Gunaydin:1983bi,Gunaydin:1984ak,Ceresole:2000jd,Bergshoeff:2004kh,Behrndt:1998ns,Behrndt:1998jd}.
We consider the model $N=2$, $D=5$ gauged supergravity, but we truncate the spectrum in order to deal with the simplest solution in $AdS_{5}$ for the present paper.

\subsection{Action}

The $N=2$, $D=5$ gauged supergravity action was constructed in \cite{D'Auria:1981kq,Gunaydin:1983bi,Gunaydin:1984ak,Ceresole:2000jd,Bergshoeff:2004kh}, coupling the pure supergravity multiplet with vector and tensor multiplets. In this paper we consider a consistent truncation of that action, in order to deal with Schwarzshild solution in $AdS_{5}$. We consider the pure supergravity multiplet, formed by the vielbein $e_{M}^{A}$, two gravitini $\psi^i_{M}$ and the graviphoton $A^{0}_{M}$, and $N-1$ vector multiplets composed by vector fields $A_{M}^{\widetilde{I}}$, gauginos $\lambda^{i\,\widetilde{I}}$ and scalar fields $q^{\widetilde{I}}$.\footnote{Index $i$ labels the two spinor fields in symplectic--Majorana representation.}

To gauge the $U\left( 1 \right)$ subgroup of $SU\left( 2 \right)$ $R$--symmetry group, we consider a linear combination of vector fields $A_{M}^{\widetilde{I}}$ and graviphoton $A_{M}$: $A_{M}=V_{I} A^{I}_{M}$, where $\{V_{I}\}$ are a set of constants and index $I$ labels the graviphoton and the $N-1$ vector fields.  The gauging procedure introduces a potential in the action which depends on the scalar $q^{\widetilde{I}}$. In order to simplify this $AdS_{5}$ model we set the potential and the scalars to constant, and the gauginos to zero. The resulting action is then
{\allowdisplaybreaks
\begin{eqnarray}\label{final}
e^{-1}\mathcal{L}&=&
%Bosonic Kinetic terms
\ft1{2} R(\omega) -\ft14a_{{I}{J}}
\widehat{F}^{{I}}_{M
N}\widehat{F}^{{J}M N}
%Fermionic Kinetic terms
-\,\ft1{2}\bar{\psi}_R \Gamma^{R M M}\mathcal{D}_M
\psi_N
%Potential
+4 g^2 \vec{P}\cdot \vec{P}
\nonumber\\&&
%Chern-Simons terms
+\,\ft{1}{6\sqrt{6}} e^{-1} \varepsilon ^{MNLRS } {\cal C}_{I J K} A_M^I \left[ F_{N L }^J F_{R S }^K + f_{FG}{}^J A_N^F A_L ^G \left(- \ft12 g F_{R S }^K + \ft1{10} g^2 f_{H G}{}^K A_R ^H A_S^G \right)\right] \nonumber\\
& & -\, \ft{1}{8} e^{-1} \varepsilon^{M N L R S}
\Omega_{I'J'} t_{IK} {}^{I'} t_{FG}{}^{J'} A_{M}^I A_N^F A_L^G
\left(-\ft 12 g  F_{R S}^K + \ft{1}{10} g^2 f_{H G}{}^K A_R^H
A_S^G \right)\nonumber\\&&
%Quadratic fermions
-\ft{\sqrt{6}}{16 }\,{\rm i}
h_{I}
F^{CD I}\bar{\psi}^A\Gamma_{ABCD}\psi^B
 +
 g\sqrt{\ft38}\,{\rm i}
P_{ij}\bar{\psi}_A^i\Gamma^{AB}\psi_B^j
%Quartic fermions
 +\ft{1}{8}\bar{\psi}_A\Gamma_B\psi^B\bar{\psi}^A\Gamma_C\psi^C\nonumber\\&&
-\,\ft{1}{16}\bar{\psi}_A\Gamma_B\psi_C\bar{\psi}^A\Gamma^C\psi^B
-\ft{1}{32}\bar{\psi}_A\Gamma_B\psi_C\bar{\psi}^A\Gamma^B\psi^C
+\ft{1}{32}\bar{\psi}_A\psi_B\bar{\psi}_C\Gamma^{ABCD}\psi_D
.
\end{eqnarray}
}
where $g$ is the $U(1)$ coupling constant. Indices $\{F,\dots,K\}$ are the special geometry ones, $\{L,M,N,\dots\}$ are the curved bulk indices and $\{A,\dots, D\}$ labels flat bulk directions. The quantities $\Omega_{IJ}$, $ {\cal C}_{I J K}$, $t_{IJ}{}^{K}$, $\vec{P}$, $h_{I}$ are related to special geometry (see \cite{Gunaydin:1984ak,Ceresole:2000jd,Bergshoeff:2004kh}). Notice that when the $i$ spinorial indices are omitted, northwest-southeast contraction is understood, e.g. $\bar{\psi}_C\psi_D = \bar{\psi}_C^{i}\psi_{i\, D} $.
We define the supercovariant field strengths $\widehat F^I_{AB}$ such that
\begin{eqnarray}
&&\widehat F^I_{AB} =F_{AB}^I-\bar{\psi}_{[A}\Gamma_{B]}\psi^I
+
\frac{\sqrt{6}}{4}\,{\rm i} \bar{\psi}_A\psi_B
h^I
,\nonumber\\
&&
F_{MN}^I \equiv  2
\partial_{[M} A_{N]}^I + g f_{JK}{}^I A_\mu^J A_\nu^K
\ .
\end{eqnarray}
We define also $\vec P \equiv  h^I \vec P_I$.
The covariant derivative reads
\begin{eqnarray}
\mathcal{D}_M \psi_{N} ^i&=&\left(\partial_M +\ft14
\omega_M{}^{AB}\Gamma_{AB}\right)\psi_{N }^i-g A_M^IP_{I}{}^{ij}\psi_{N
j} .
\end{eqnarray}
This action admits the following $N=2$ supersymmetry:
\begin{eqnarray}
\delta  e_M{}^A &=& \ft12 \bar\e \Gamma^A \psi_M ,\nonumber\\
\delta \psi_M^i&=&D_\mu(\hat\omega)\e^i + \ft{{\rm i}} {4\sqrt6}
h_{ I} {\widehat{F}}^{ I N R}
 ({\Gamma}_{MN R} - 4 g_{M N} {\Gamma}_R) \e^i  - \ft{1}{\sqrt6} {\rm i} g P^{ij}\Gamma_M  \e_j
,\nonumber\\
\delta  A_M^I&=&- \ft{\sqrt6}{4 } \,{\rm i} h^{ I} \bar\e \psi_M
  .
\end{eqnarray}
We also denoted
\begin{eqnarray}
D_M(\hat\omega) \e^i &=& {\mathcal D}_M(\hat\omega) \e^i  - g A_M^I P_I^{ij}\e_j,
\end{eqnarray}
where $\hat\omega$ indicates the spin connection defined through vielbein postulate, as we will see in the forthcoming sessions.

\subsection{Spinors Relations}

For our purpose, we find convenient to work with Dirac spinors instead of symplectic--Majorana.\footnote{Dirac spinors are also used in \cite{Behrndt:1998ns,Behrndt:1998jd} while symplectic--Majorana ones are present in \cite{D'Auria:1981kq,Gunaydin:1983bi,Gunaydin:1984ak,Ceresole:2000jd,Bergshoeff:2004kh}. }
Therefore we dedicate the present subsection to illustrate and remind the reader the translation table.

For $5$ dimensions SM spinors $\lambda^{i}$ with $i=\left\{ 1,2 \right\}$, the complex conjugate is defined through
\begin{align}
	(\lambda^{i})^{*} = &\  C \Gamma_{0} \lambda^{i}\ ,
	\label{SMspinor1}
\end{align}
the bar is the Majorana bar
\begin{align}
	\bar \lambda^{i} =&\ (\lambda^{i})^{T}\mathcal{C} \ ,
	\label{SMspinor2}
\end{align}
where $\mathcal{C}$ is the charge conjugation matrix satisfying
\begin{align}
	\mathcal{C}^{T}= -\mathcal{C}\ ,&
	&
	\mathcal{C}^{*}= -\mathcal{C}\ ,&
	&
	\mathcal{C}^{2}=\mathcal{C}^{\dagger}C=I\ ,&
	\nonumber\\
	&
	\left( \mathcal{C}\Gamma_{M} \right)^{T}=-C\Gamma_{M}\ ,&
	&
	\Gamma_{M}^{T}=\mathcal{C}\Gamma_{M}\mathcal{C}^{-1}\ .&
	\label{SMspinorC}
\end{align}
Thus, the following expressions are real
\begin{align}
	i \bar\lambda^{i}\psi_{i} \ ,
	\qquad\qquad
	\bar\lambda^{i}\Gamma_{M}\psi_{i} \ .
	\label{SMspinor3}
\end{align}
Notice that the index $i$ is raised and lowered by the antisymmetric tensor $\varepsilon_{ij}$.

For our purpose, we need Dirac spinors $\epsilon$ and the bar represents the Dirac adjoint
\begin{align}
	\bar\epsilon = \epsilon^{\dagger}\Gamma_{0}\ .
	\label{Dspinor1}
\end{align}
It is possible to construct one Dirac spinor from two SM: one has $\epsilon=\lambda_{1}+i \lambda_{2}$. For consistency then we have $\bar\epsilon=\bar\lambda_{1}-i\bar\lambda_{2}$.

Using the above relations we express the quantities (\ref{SMspinor3}) in terms of Dirac spinors
\begin{align}
	i \bar\lambda^{i}\psi_{i}
	=
	\textrm{Re}\left( \bar\epsilon \psi \right)\ ,
	\qquad\qquad
	\bar\lambda^{i}\Gamma_{M}\psi_{i}
	=
	\textrm{Re}\left( -i \bar\epsilon\Gamma_{M} \psi \right)\ ,
	\label{Dspinor2}
\end{align}
where $\textrm{Re}(x)$ denotes the real part of $x$.

\subsection{Susy Transformations}

The supersymmetry transformations~(\ref{AlgSusyTransfv1}) for $N=2$, $D=5$ gauged supergravity written with Dirac spinors are
\begin{align}
	\delta_{\epsilon} e_{M}^{A}
		= &
		-\frac{1}{2} {\textrm{Re}}\left( i \bar\epsilon \Gamma^{A} \delta\psi_{M} \right)
	\ ,
	\nonumber\\
	\delta_{\epsilon} g_{MN}
		= &
	-\frac{1}{2} {\textrm{Re}}\left( i  \bar\epsilon \Gamma_{\left( M \right.} \delta\psi_{\left. N \right)} \right)
	\ ,
	\nonumber\\
	\delta_{\epsilon} \psi_{M}
		= &
		\mathcal{D}_{M}\left( \hat\omega \right)\epsilon
	+
	\frac{i}{4\sqrt{6}} e^{a}_{M} h_{I} \hat F^{I\,BC}
	\left(
		\Gamma_{ABC}-4\eta_{AB}\Gamma_{C}
	\right) \epsilon
	\ ,
	\nonumber\\
	\delta_{\epsilon}A_{M}^{I}
		= &
	-\frac{\sqrt{6}}{4}{\textrm{Re}}
	\left(  \bar\epsilon \psi_{M}h^{I}  \right)
	\ ,
	\label{AlgSusyTransf}
\end{align}
where
\begin{align}
	\hat F_{AB}^{I}
	= &
	F_{AB}^{I}
	+
	\frac{\sqrt{6}}{4}\bar\psi_{\left[ A \right.} \psi_{\left. B \right]}
	h^{I}
	\ ,
	\nonumber\\
	\mathcal{D}_{M}\left( \hat\omega \right)
	= &
	D_{M}\left( \hat\omega \right)	
	-g A_{M}^{I}P_{I}
	\ ,
	\nonumber\\
	D_{M}\left( \hat\omega \right)
	= &
	\partial_{M} +\frac{1}{4} \hat\omega_{M}^{AB} \Gamma_{AB} -
	\frac{i}{\sqrt{6}}g P \Gamma_{M} %+\frac{1}{2} e_{\mu}^{a}\Gamma_{a}
	\ .
	\label{AlgDef1}
\end{align}
In order to compare this with the $AdS$ covariant derivative
\begin{equation}
	D_{M}\left( \hat\omega \right)
	=
	\partial_{M} +\frac{1}{4} \hat\omega_{M}^{AB} \Gamma_{AB}
	+\frac{1}{2} e_{M}^{A}\Gamma_{A}
	\ ,
	\label{AlgCovDerTipical}
\end{equation}
we set
\begin{align}
	g P
	= &
	\frac{i}{2}\sqrt{6}
	\ .
	\label{AlgDefKillPrep}
\end{align}
From the special geometry construction, $h^{I}$ satisfies
\begin{equation}
	h_{I}h^{I} = 1
	\ ,
	\label{AlgSpecGeom}
\end{equation}
then, in our particular case, where the gauge fields are generated only from susy transformation (\ref{AlgSusyTransf}) while the zero--order is zero, we define the gauge field as
\begin{equation}
	A_{M}^{I} = A_{M} h^{I}
	\ .
	\label{AlgRedefA}
\end{equation}
Doing so, all the indices $I$ and the quantity $h^{I}$ disappear from the equations. Moreover, using  eq.~(\ref{AlgDefKillPrep}), the $A$--part in the covariant derivative becomes
\begin{align}
	-g A_{M}^{I}P_{I}
	=
	-\frac{i}{2}\sqrt{6} \, A_{M}
	\ .
	\label{AlgNewCovDerPiece}
\end{align}
Finally, the simplified susy transformations now read
\begin{align}
	\delta_{\epsilon} e_{M}^{A}
		= &
	-\frac{1}{2} {\textrm{Re}}\left(   i  \bar\epsilon \Gamma^{A} \delta\psi_{M} \right)
	\ ,
	\nonumber\\
	\delta_{\epsilon} g_{MN}
		= &
	-\frac{1}{2} {\textrm{Re}}\left( i \bar\epsilon \Gamma_{\left( M \right.} \delta\psi_{\left.N \right)} \right)
	\ ,
	\nonumber\\
	\delta_{\epsilon} \psi_{M}
		= &
		\mathcal{D}_{M}\left( \hat\omega \right)\epsilon
	+
	\frac{i}{4\sqrt{6}} e^{A}_{M}  \hat F^{BC}
	\left(
		\Gamma_{ABC}-4\eta_{AB}\Gamma_{C}
	\right) \epsilon
	\ ,
	\nonumber\\
	\delta_{\epsilon}A_{M}
		= &
	-\frac{\sqrt{6}}{4}{\textrm{Re}}
	\left(  \bar\epsilon \psi_{M} \right)
	\ ,
	\label{AlgSusyTransfv1}
\end{align}
where
\begin{align}
	\hat F_{AB}
	= &
	F_{AB}
	+
	\frac{\sqrt{6}}{4}\bar\psi_{\left[ A \right.} \psi_{\left. B \right]}
	\ ,
	\nonumber\\
	\mathcal{D}_{\mu}\left( \hat\omega \right)
	= &
	D_{\mu}\left( \hat\omega \right)	
	-\frac{i}{2}\sqrt{6} \, A_{M}\,
	\ ,
	\nonumber\\
	D_{M}\left( \hat\omega \right)
	= &
	\partial_{M} +\frac{1}{4} \hat\omega_{M}^{AB} \Gamma_{AB}
	+\frac{1}{2} e_{M}^{A}\Gamma_{A}
	\ .
	\label{AlgDef1v1}
\end{align}

As last remark, notice that torsion is not zero:
\begin{equation}
	\dd e^{A} +\omega^{A}_{\phantom{A}B}\wedge e^{B}
	=
	\frac{i}{4} \bar\psi \Gamma^{A} \psi
	\ ,
	\label{AlgTorsion}
\end{equation}
then, the spin connection $\hat\omega$ is written in terms of both vielbein and gravitino bilinears. Moreover, the abelian field strength reads
\begin{align}
	F_{MN}
	= &
	D_{M} A_{N} - D_{N} A_{M}
	=
	\partial_{M} A_{N} - \partial_{N} A_{M}
	+i \frac{1}{4} \bar\psi_{\left[ M \right.}\Gamma^{A}\psi_{\left. N \right]} A_{A}
	\ .
	\label{AlgAbelianFmunu}
\end{align}
We are left with the vielbeins, the gauge field and the Rarita-Schwinger (RS) field, which form the $N=2$, $D=5$ pure supergravity. Now, we can truncate to the bosonic sector and we consider a Schwarzschild--type solution which is asymptotically $AdS$.
Of course there are also more intricated solutions with non--constant scalar fields or gauge fields, but we do not take these cases into account in the present work.

\subsection{Background Setup}

We choose a $AdS_{5}$ solution of pure Einstein gravity as background
\begin{equation}
\dd s^{2} =
- r^{2} \dd t^{2}
+ \frac{1}{r^{2}} \dd r^{2}
+ r^{2} \sum_{i = 1}^{3} \dd x_{i}^{2}
\ ,
\qquad
A_{M}=0
\ ,
\qquad
\psi_{M}=0
\ ,
\label{AdSmetric0}
\end{equation}
where the metric is given in the Poicar\'e patch.
Notice that in this initial set up the gauge field and the Rarita--Schwinger fields are set to zero \cite{Burrington:2004hf} and  $AdS_{5}$ radius is set to $1$. The associated non-zero vielbein components are
\begin{align}\label{AdSvielbein0}
	e^{0}_{t} 	
&=
	r\ ,
&
	e^{1}_{r} 	
&=
	\frac{1}{r}\ ,
&
	e^{a}_{i} 	
&=
	r\delta_{i}^{a}
\ ;
\end{align}
while the non-zero spin connection components are
\begin{align}\label{AdSspincon0}
	\omega^{01}_{t} 	
&=
	r\ ,
&
	\omega^{a1}_{i} 	
&=
	r\delta_{i}^{a}
\ .
\end{align}
Notice that we will use capital latin letters to indicate bulk directions (i.e. $M,N$ run from $0$ to $4$) leaving greek alphabet to boundary ones (i.e. $\mu,\nu$ run from $0$ to $3$) furthermore $\left\{ t,r,i \right\}$ are {\it curved} indices and $\left\{ 0,1,a \right\}$ represent {\it flat} ones.

In presence of a uncharged, irrotational black hole eq.~(\ref{AdSmetric0}) becomes
\begin{equation}
\dd s^{2} =
- \left( r^{2}+\frac{\mu}{r^{2}} \right) \dd t^{2}
+ \frac{1}{ r^{2}+\frac{\mu}{r^{2}} } \dd r^{2}
+ r^{2} \sum_{i = 1}^{3} \dd x_{i}^{2}
\ ,
\label{AdSmetricBH}
\end{equation}
in this case the non-zero vielbein components are
\begin{align}\label{AdSvielbeinBH}
	e^{0}_{t} 	
&=
	\sqrt{ r^{2}+\frac{\mu}{r^{2}}}\ ,
&
	e^{1}_{r} 	
&=
	\frac{1}{\sqrt{ r^{2}+\frac{\mu}{r^{2}}}}\ ,
&
	e^{a}_{i} 	
&=
	r\delta_{i}^{a}
\ ;
\end{align}
and the non-zero spin connection components are
\begin{align}\label{AdSspinconBH}
	\omega^{01}_{t}
&=
	r-\frac{\mu}{r^{3}}\ ,
&
	\omega^{a1}_{i} 	
&=
	\sqrt{ r^{2}+\frac{\mu}{r^{2}}}\,\delta_{i}^{a}
\ .
\end{align}
Convenient coordinates are the Eddington-Finkelstein ones. They are defined through the following change of variables:
\begin{equation}
t = v + \frac{1}{r}
\ ,
\label{EddFincoord0}
\end{equation}
thus we get
\begin{equation}
\dd^{2} g =
- r^{2} \dd v^{2}
+ 2 \dd r \dd v
+ r^{2} \sum_{i = 1}^{3} \dd x_{i}^{2}
\ .
\label{AdSmetricEF0}
\end{equation}
In this case the non-zero vielbein components are
\begin{align}
	e^{0}_{v} 	
&=
	r\ ,
&
	e^{0}_{r} 	
&=
	-\frac{1}{r}\ ,
&
	e^{1}_{r} 	
&=
	\frac{1}{r}\ ,
&
	e^{a}_{i} 	
&=
	r\delta_{i}^{a}
\ ;
	\label{AdSvielbeinBHef}
\end{align}
while the non-zero components of spin connection are
\begin{align}\label{AdSspinconBHef}
	\omega^{01}_{v}
&=
	r\ ,
&
	\omega^{01}_{r} 	
&=
	-\frac{1}{r}
&
	\omega^{a1}_{i} 	
&=
	r\,\delta_{i}^{a}
\ .
\end{align}
Eq.~(\ref{AdSmetricBH}) in this coordinates system is
\begin{equation}
\dd s^{2} =
- \left( r^{2}+\frac{\mu}{r^{2}} \right) \dd v^{2}
+ 2 \dd r \dd v
+ r^{2} \sum_{i = 1}^{3} \dd x_{i}^{2}
\ ,
\label{AdSmetricEF_BH}
\end{equation}
where we used the following change of coordinates
\begin{equation}
t = v - \int \frac{1}{ r^{2}+\frac{\mu}{r^{2}}}\,\dd r
\ .
\label{EddFincoordBH}
\end{equation}

\section{Killing Vectors for $AdS_{5}$}

The basis \cite{Bhattacharyya:2008jc,Rangamani:2009xk} for deriving the boundary equations of motion is the analysis of the isometries of $AdS$ space.
On a second step one can evaluate which of those isometries are preserved by the black hole solutions and in terms of the broken isometries one can build
 local transformations, where the parameters are replaced by local expansion on the boundary coordinates.

Even though we are interested here only in the fermionic wigs, we present the form of bosonic Killing vectors.
That will turn to be useful in the forthcoming analysis.

The Killing vectors for metric~(\ref{AdSmetric0}) read
\begin{align}
K^{t} = &
-
\left( \frac{t^{2}}{2} + \frac{1}{2r^{2}} \right) c
-
t\left( x_{j} e_{j} + e \right)
-
\frac{1}{2} x_{j}x_{j} c
+
d_{j} x_{j}
+
d
\ ,\nonumber\\
K^{r} = &
r t c+ r\left( x_{j}e_{j} + e \right)
\ ,\nonumber\\
K^{i} = &
\left(\frac{1}{2r^{2}} - \frac{t^{2}}{2} \right) e^{i}
-
t x^{i} c
+
t d^{i}
+
\frac{1}{2}x_{j}x^{j} e^{i}
-
x^{i} x_{j} e^{j}
-
x^{i} e
+
w^{ij} x_{j}
+
h^{i}
\ ,
\label{KV0}
\end{align}
where the $15$ infinitesimal parameters are interpreted as follows: $\left\{ d_{i} \right\}$ are the boundary boost parameters,
$\left\{ d, h_{i} \right\}$ represent translations in $\left\{t, x^{i}\right\}$ directions,
$e$ is the dilatation,
$\left\{ c, e_{i} \right\}$ are associated to conformal transformations
and $\left\{w_{ij}\right\}$ is the antisymmetric tensor responsible of the $3$ rotations in $\left\{x_{i}\right\}$.

The variation of the black hole metric in the Eddington-Finkelstein coordinates (\ref{AdSmetricEF_BH}), generated by these Killing vectors with all the conformal parameters set to zero reads
\begin{align}
	\dd s^{2} = &
	2 \dd v\,\dd r- h^{2}\left(r\right) \dd v^{2}
	+r^{2}\dd x_{i}\dd\, x^{i}
	+\nonumber\\
	&
	-2b_{i}\left( 1-\frac{r^{2}}{h^{2}\left(r\right)} \right) \dd x^{i}\, \dd r
	-2b_{i}\left( r^{2}-h^{2}\left(r\right) \right)\dd x^{i}\,\dd v
	+4 \mu \frac{b}{r^{2}} \dd v^{2}
	\ ,
	\label{MinwOurBBB}
\end{align}
where $h\left(r\right)=\sqrt{r^{2}+\frac{\mu}{r^{2}}}$.
In the work \cite{Bhattacharyya:2008jc,Rangamani:2009xk} it has been chosen a different frame, and that is achieved by setting $\mu=-1$ and by a change of coordinate
generated by the following vectors
\begin{align}
	\xi^{i} = &
	\int \frac{f^{i}\left( r \right)}{r^{2}}\dd r + \hat w^{i}_{j}x^{j}+\hat{d}^{i}
	\ ,\nonumber\\
	\xi^{r} = & \xi^{v} = 0 \ ,
	\label{diff1}
\end{align}
where $f^{i}\left( r \right)= 2 b_{i}\frac{r^{2}}{h^{2}\left(r\right)}\ $, $\hat w^{i}_{j}$ is an antisymmetric matrix and $\hat{d}^{i}$ is a constant. We get
\begin{align}
	\dd s^{2} = &
	2 \dd v\,\dd r- h^{2}\left(r\right) \dd v^{2}
	+r^{2}\dd x_{i}\dd\, x^{i}
	+\nonumber\\
	&
	-2b_{i} \dd x^{i}\, \dd r
	-2b_{i}\left( r^{2}-h^{2}\left(r\right) \right)\dd x^{i}\,\dd v
	-4 \frac{b}{r^{2}} \dd v^{2}
	\ .
	\label{MinwBBB}
\end{align}

\section{Killing Spinors for $AdS_{5}$}

Here we compute $AdS$ Killing spinors. We found that there are two independent solutions.
These are obtained by first factorizing the Dirac spinors into a $2d$ spinor and a $3d$ spinor in their irreducible representations.

Notice that, since we are interested into the complete solution -- namely all powers of fermions -- we have to deal with the fermionic nature of the spinor fields.
Therefore, factorizing the spinors into a product of spinors in lower dimensions, we have to declare the statistic of each part.
As a matter of fact, we saw that the map between the original fermion $\epsilon$ and its decomposition $\varepsilon\otimes\eta$
 spoils the correct number of degrees of freedom only if all possible choices are taken into account.
Namely, we have to choose first $\varepsilon$ to be anticommuting and $\eta$ commuting and subsequently $\varepsilon$ commuting and $\eta$ anticommuting:
\begin{align}
	\epsilon = \varepsilon|_{A} \otimes \eta|_{C} + \varepsilon|_{C} \otimes \eta|_{A}\ .
	\label{KScomment}
\end{align}
The generalization to an arbitrary number of dimensions is straightforward.
As we will see, in the present case $\varepsilon$ has only one degree of freedom. This allows us to consider
just $\epsilon=\varepsilon|_{C} \otimes \eta|_{A}$.
In the forthcoming we will drop indices $A,C$.

The Killing spinor equation for $AdS$ reads
\begin{equation}
\left(
\partial_{M}
+\frac{1}{4}\omega_{M}^{ab}\Gamma_{ab}
+
\frac{1}{2}e_{M}^{a}\Gamma_{a}
 \right)\epsilon = 0
\ .
\label{KSeq0}
\end{equation}
with $\Gamma_{ab}=\frac{1}{2}\left( \Gamma_{a}\Gamma_{b}-\Gamma_{b}\Gamma_{a} \right)$. In components we have
\begin{align}
	&
	\partial_{t}\epsilon + \frac{r}{2}\Gamma_{0}\left( \Gamma_{1}+\one \right)\epsilon=0\ ,
	\nonumber\\
	&
	\partial_{r}\epsilon+\frac{1}{2r}\Gamma_{1}\epsilon=0\ ,
	\nonumber\\
	&
	\partial_{i}\epsilon + \frac{r}{2} \Gamma_{i}\left( \Gamma_{1}+\one \right)\epsilon=0\ .
	\label{ksEQ1}
\end{align}
We can divide the $5$ dimensional space in two parts: $\left\{ t,r \right\}$ and $\left\{ x^{i} \right\}$ using the following gamma matrices parametrization
\begin{align}
	\Gamma_{0} & = i\sigma_{2}\otimes\hat\sigma_{0}
\ ,
&\Gamma_{1} & = \sigma_{1}\otimes\hat\sigma_{0}
\ ,
&\Gamma_{a} & = \sigma_{3}\otimes\hat\sigma_{a}
\ ,
\label{Gamma}
\end{align}
where $\sigma_{0}$ is the identity matrix in $2d$. Hatted matrices refer to ${x^{i}}$ space. In this way, the solution of eq.~(\ref{KSeq0}) is
\begin{eqnarray}
\epsilon
=
\left( \frac{1}{\sqrt{r}}- t \sqrt{r} \sigma_{3} \right) \varepsilon_{0}\otimes \eta_{1}
-
\sqrt{r}\sigma_{3}\varepsilon_{0}\otimes\eta_{2}
\ ,
\label{KSsolution1}
\end{eqnarray}
where
\begin{align}
\eta_{2} = &
x^{k} \hat\sigma_{k} \eta_{1}+\eta_{0}\ ,
\label{KSsolution11}
\end{align}
and $\eta_{1}\,,\,\eta_{0}$ are $2$--dimensional complex spinors (and so contain $8$ real dof's) while $\varepsilon_{0}$ is a real $2$--dimensional spinor with only one dof. The total number of degrees of freedom is then $1\times8$. The solution~(\ref{KSsolution1}) can also be written as
\begin{align}
	\epsilon
	=
	\frac{1}{\sqrt{r}}\sigma_{0}\otimes\hat\sigma_{0}\, \varepsilon_{0}\otimes\eta_{1}
	-\sqrt{r}\sigma_{3}\otimes \left( t\hat\sigma_{0}+x^{i}\hat\sigma_{i} \right)\, \varepsilon_{0}\otimes\eta_{1}
	-\sqrt{r}\sigma_{3}\otimes\hat\sigma_{0}\, \varepsilon_{0}\otimes\eta_{0}
	\ .
	\label{KSsolution2}
\end{align}
Notice that $\bar\epsilon\,\Gamma^{M}\epsilon$ reproduces the Killing vectors (\ref{KV0}) as expected.

\section{Algorithms}\label{ALG}

To build the BH wigs, we use the following algorithm. We expand in powers of fermionic bilinears.
Notice that we could have performed a finite supersymmetry transformation,
however it turns out to be more convenient dealing with an iterative procedure due to the anticommuting character of fermions.

\subsection{Generalities for Algorithms}\label{generalities}

The algorithms are based on the perturbative expansions in fermionic bilinears (for the bosonic quantities) or spinors (for fermionic ones). Then, every quantity is labelled by an integer index between square brakets $\left[ N = 1\cdots\right]$ denoting the perturbative order.

More in detail:
\begin{itemize}
	\item $e_{M}^{\left[ 1 \right] A}$, $e^{\left[ 1 \right]M}_{A}$ and $\omega_{M}^{\left[ 1 \right]A B}$ contain zero bilinears;
	\item $e_{M}^{\left[ 2 \right] A}$, $e^{\left[ 2 \right]M}_{A}$ and $\omega_{M}^{\left[ 2 \right]A B}$ contain one bilinear;
	\item $e_{M}^{\left[ N \right] A}$, $e^{\left[ N \right]M}_{A}$ and $\omega_{M}^{\left[ N \right]A B}$ contain $N-1$ bilinears;
\end{itemize}
but
\begin{itemize}
	\item $\delta^{\left[ 1 \right]} g_{M N}$, $\delta^{\left[ 1 \right]} A_{M}$ contain one bilinear;
	\item $\delta^{\left[ N \right]} g_{M N}$, $\delta^{\left[ N \right]} A_{M}$ contain $N$ bilinear;
\end{itemize}
and
\begin{itemize}
	\item $\delta^{\left[ 1 \right]} \psi_{M}$ contains one spinor ($1/2$ bilinear);
	\item $\delta^{\left[ N \right]} \psi_{M}$ contains $2N-1$ spinors ($N-1/2$ bilinears).
\end{itemize}

\subsection{Inverse Vielbein $e_{A}^{M}$}

To compute the inverse vielbein $e_{A}^{M}$, we use the definition
\begin{equation}
	e_{M}^{A} e_{B}^{M} = \delta^{A}_{B}
	\ ,
	\label{AlgEI0}
\end{equation}
expanding the vielbeins we get
\begin{equation}
	\left(
	e_{M}^{\left[ 1 \right]A}	
	+
	e_{M}^{\left[ 2 \right]A}
	+
	e_{M}^{\left[ 3 \right]A}
	\right)
	\left(
	e_{B}^{\left[ 1 \right]M} 	
	+
	e_{B}^{\left[ 2 \right]M} 	
	+
	e_{B}^{\left[ 3 \right]M} 	
	\right)
	= \delta^{A}_{B}
	\ .
	\label{AlgEI1}
\end{equation}
We then obtain one equation for each perturbative order
\begin{align}
	\delta^{A}_{B} = & e_{M}^{\left[ 1 \right]A} e_{B}^{\left[ 1 \right]M}
	\ ,
	\nonumber\\
	0 = &
	e_{M}^{\left[ 1 \right]A} e_{B}^{\left[ 2 \right]M}
	+
	e_{M}^{\left[ 2 \right]A} e_{B}^{\left[ 1 \right]M}
	\ ,
	\nonumber\\
	0 = &
	e_{M}^{\left[ 1 \right]A} e_{B}^{\left[ 3 \right]M}
	+
	e_{M}^{\left[ 2 \right]A} e_{B}^{\left[ 2 \right]M}
	+
	e_{M}^{\left[ 3 \right]A} e_{B}^{\left[ 1 \right]M}
	\ .
	 \label{AlgEI2}
\end{align}
The first one is solved as usual by inverting the vielbein $e_{M}^{A}$.
The other equations are solved by
\begin{align}
	e^{\left[ 2 \right]M}_{B}
	= &
	- e^{\left[ 1 \right]M}_{A}
		\left[
			e^{\left[ 2 \right] A}_{ R}
			e^{\left[ 1 \right] R}_{B}
		\right]
	\ ,\nonumber\\
	e^{\left[ 3 \right]M}_{B}
	= &
	- e^{\left[ 1 \right]M}_{A}
		\left[
			e^{\left[ 2 \right] A}_{ R}
			e^{\left[ 2 \right] R}_{B}
			+
			e^{\left[ 3 \right] A}_{ R}
			e^{\left[ 1 \right] R}_{B}
		\right]
	\ .
	\label{AlgEI3}
\end{align}
In general we have, for $N > 1$
\begin{align}
	e^{\left[ N \right]M}_{B}
	= &
	- e^{\left[ 1 \right]M}_{A}
	V^{\left[ N \right] A}_{\phantom{\left[ N \right] }B}
	\ ,\nonumber\\
	V^{\left[ N \right] A}_{\phantom{\left[ N \right] }B}
	= &
	\sum_{p=1}^{N-1}
	e^{\left[ p+1 \right] A}_{ R}
	e^{\left[ N-p \right] R}_{B} \ .
	\label{AlgEIfin}
\end{align}

\subsection{Spin Connection $\omega_{M}^{AB}$}

The spin connection $\omega_{M}^{A B}$ is defined by the vielbein postulate (\ref{AlgTorsion})
\begin{equation}
	\dd e^{A} + \omega_{\phantom{a}B}^{A}\wedge e^{B} = \frac{i}{4} \bar\psi \Gamma^{A} \psi\ .
	\label{AlgSpinC0}
\end{equation}
Extracting the $1$--form basis $\left\{ \dd x^{M} \right\}$, it becomes
\begin{eqnarray}
	\partial_{\left[ M \right.} e^{A}_{\left.  N \right]}
	+
	\omega^{A B}_{\left[ M \right.}\,\eta_{B C}\, e^{C}_{\left.  N \right]}
	=
	\frac{i}{4}
	\bar\psi_{\left[ M \right.} \Gamma^{A} \psi_{\left.  N \right]} \ .
	\label{AlgSpinC1}
\end{eqnarray}
As in the case of inverse vielbein, we expand in perturbative order.  We obtain the following result
\begin{align}
	\omega_{M}^{\left[ N \right]DC}
	= &
	e^{\left[ 1 \right]}_{M\,A}\left[
		\Omega^{\left[ N \right]\,D C\,,A}
		-
		\Omega^{\left[ N \right]\,C A\,,D}
		-
		\Omega^{\left[ N \right]\,A D\,,C}
		\right]
	\ ,\nonumber\\
	\Omega^{\left[ N \right]\,D C\,,A}
	= &
	e^{\left[ 1 \right]  N \left[ D \right.} e^{\left[ 1 \right] M \left. C \right]}
		\left[
			\partial_{\left[ M \right.} e^{\left[ N \right]\,A}_{\left. N \right]}
			+
			\sum_{p=1}^{N-1}\omega^{\left[ N-p \right]\,A B}_{\left[ M \right.}\,\eta_{B C}\,e^{\left[ p+1 \right]\,C}_{\left. N \right]}
			-
			\frac{i}{4} \sum_{p=1}^{N-1}\eta^{A B}\, \bar\psi^{\left[ p \right]}_{\left[ M \right.}\Gamma_{B}\psi^{\left[ N-p \right]}_{\left. N \right]}
		\right]
	\label{AlgSpinCfin}
\end{align}

\subsection{Gravitino}

Using definitions (\ref{AlgSusyTransfv1}), (\ref{AlgDef1v1}),  and (\ref{AlgAbelianFmunu}) we get\begin{align}
	\delta_{\epsilon}\psi_{M}
	= &
	\left(
	\partial_{M}
	+
	\frac{1}{4} \hat\omega_{M}^{A B} \Gamma_{A B}
	+
	\frac{1}{2} e_{M}^{A}\Gamma_{A}
	-
	\frac{i}{2}\sqrt{6} \, A_{M}
	\right)\epsilon
	+\nonumber\\
	& +
	\frac{i}{4\sqrt{6}}\, e^{A}_{M} \,
	\left( \Gamma_{A B C} - 4 \eta_{A B} \Gamma_{C} \right) \epsilon
	\, \eta^{B B'} \, \eta^{C C'}
	\left[
		e_{B'}^{ R}\, e_{C'}^{ S}
	\right] \times
	\nonumber\\
	& \times
	\left[
		\partial_{ R} A_{ S} - \partial_{ S} A_{ R}
		+
		\frac{i}{4}
			\bar\psi_{\left[  R \right.}
			\,\Gamma_{A}\,
			\psi_{\left.  S \right]}
			\,\eta^{A A'}\,A_{A'}
		+
		\frac{\sqrt{6}}{4}
		\bar\psi_{\left[  R \right.} \psi_{\left.  S \right]}
	\right]
	\ .
	\label{AlgGrav1}
\end{align}
In order to compute the gravitino variation $\psi^{\left[ N \right]}_{M}$ order by order we separate the expression above in different pieces
\begin{itemize}
	\item $
	\mathcal{D}^{\left[ N \right]}_{M}\left( \hat\omega \right)\epsilon = \left(
		\partial^{\left[ N \right]}_{M}
	+
	\frac{1}{4} \hat\omega_{M}^{\left[ N \right] A B} \Gamma_{A B}
	+
	\frac{1}{2} e_{M}^{\left[ N \right]A}\Gamma_{A}
	-
	\frac{i}{2}\sqrt{6} \, A_{M}^{\left[ N-1 \right]}
	\right)\epsilon$:
	this part contains ``$2N-1$ spinors'' (short way to say $N-1$ bilinears and one spinor).
	Notice that $\partial^{\left[ N \right]}_{M}\epsilon$ is simply zero for $N>1$;

	\item $%\left( A^{\left[ N \right]} \right)^{a}_{M} =
		e^{\left[ N \right] A}_{M} $: contains $N-1$ bilinears ($2\left( N-1 \right)$ spinors);
	
\item $\left( B \right)_{A B C} = 	\left( \Gamma_{A B C} - 4 \eta_{A B} \Gamma_{C} \right) \epsilon $: this term contains always only one spinor $\epsilon$;

	\item $\left( C^{\left[ N \right]}  \right)^{ R S}_{B'C'} =\left[ e_{B'}^{ R}\, e_{C'}^{ S} \right]^{\left[ N \right]}$;

	\item $\left( D_{0}^{\left[ N \right]}  \right)_{ R S} = \partial_{ R} A_{ S}^{\left[ N \right]}  - \partial_{ S} A_{ R}^{\left[ N \right]}   $;

	\item $ \left( D_{1}^{\left[ N \right]}  \right)_{ R S\, A} =-i\left[ \bar\psi_{\left[  R \right.}
			\,\Gamma_{A}\,
			\psi_{\left.  S \right]} \right]^{\left[ N \right]} $;

		\item $ \left( D_{2}^{\left[ N \right]}  \right)_{ R S} = \left[ \left( D_{1} \right)_{ R S\, A}
			\,\eta^{A A'}\,A_{A'}  \right]^{\left[ N \right]}$;

		\item $\left( D_{3}^{\left[ N \right]} \right)_{ R S} =-i\left[  \bar\psi_{\left[  R \right.} \psi_{\left.  S \right]} \right]^{\left[ N \right]}$.
\end{itemize}
With these definitions, (\ref{AlgGrav1}) becomes
\begin{align}
	\delta_{\epsilon}^{\left[ N \right]}\psi_{M}
	= &
	\mathcal{D}^{\left[ N \right]}_{M}\left( \hat\omega \right)\epsilon
	+
	\frac{i}{4\sqrt{6}}
		\left( e^{\left[ N_{e} \right]} \right)^{A}_{M}
		\left( B \right)_{A B C}
		\left( C^{\left[ N_{C} \right]}  \right)^{ R S}_{D E}
		\, \eta^{B D} \, \eta^{C E} \times
	\nonumber\\ &
	\times
	\left[ D_{0}
	-
	\frac{1}{4} D_{2}
	+
	\frac{i\sqrt{6}}{4} D_{3}
	\right]^{\left[ N_{D} \right]} _{ R S}
	\ .
	\label{AlgFrav2}
\end{align}
To obtain the correct perturbative order $\left[ N \right]$ for $\delta_{\epsilon}^{\left[ N \right]}\psi_{M}$ the quantites  $N_{e}, N_{B}, N_{C}$ and $N_{D}$ must take the value as shown in the following table.
\begin{center}
\begin{tabular}{|c|c|c|c|c|}
\hline
$N$ & $N_{e}$ & $N_{B}$ & $N_{C}$ & $N_{D}$ \\
\hline
$\left[ 1 \right]\,\left( 1/2 \right) $ & $\left[ 1 \right]\, \left( 0 \right)$ & $\left( 1/2 \right)$ & $\left[ 1 \right]\, \left( 0 \right)$ & $0$ \\
\hline
$\left[ 2 \right]\,\left( 3/2 \right)$ & $\left[ 1 \right]\, \left( 0 \right)$ & $\left( 1/2 \right)$ & $\left[ 1 \right]\,\left( 0 \right)$ & $\left[ 1 \right] \left( 1 \right)$ \\
\hline
$\left[ 3 \right]\,\left( 5/2 \right)$ & $\left[ 1 \right]\, \left( 0 \right)$ & $\left( 1/2 \right)$ & $\left[ 1 \right]\,\left( 0 \right)$ & $\left[ 2 \right] \left( 2 \right)$ \\
 & $\left[ 2 \right]\, \left( 1 \right)$ & $\left( 1/2 \right)$ & $\left[ 1 \right]\,\left( 0 \right)$ & $\left[ 1 \right] \left( 1 \right)$ \\
 & $\left[ 1 \right]\, \left( 0 \right)$ & $\left( 1/2 \right)$ & $\left[ 2 \right]\,\left( 1 \right)$ & $\left[ 1 \right] \left( 1 \right)$ \\
\hline
$\left[ 4 \right]\,\left( 7/2 \right)$ & $\left[ 1 \right]\, \left( 0 \right)$ & $\left( 1/2 \right)$ & $\left[ 1 \right]\,\left( 0 \right)$ & $\left[ 3 \right] \left( 3 \right)$ \\
 & $\left[ 2 \right]\, \left( 1 \right)$ & $\left( 1/2 \right)$ & $\left[ 1 \right]\,\left( 0 \right)$ & $\left[ 2 \right] \left( 2 \right)$ \\
 & $\left[ 1 \right]\, \left( 0 \right)$ & $\left( 1/2 \right)$ & $\left[ 2 \right]\,\left( 1 \right)$ & $\left[ 2 \right] \left( 2 \right)$ \\
 & $\left[ 2 \right]\, \left( 1 \right)$ & $\left( 1/2 \right)$ & $\left[ 2 \right]\,\left( 1 \right)$ & $\left[ 1 \right] \left( 1 \right)$ \\
 & $\left[ 3 \right]\, \left( 2 \right)$ & $\left( 1/2 \right)$ & $\left[ 1 \right]\,\left( 0 \right)$ & $\left[ 1 \right] \left( 1 \right)$ \\
 & $\left[ 1 \right]\, \left( 0 \right)$ & $\left( 1/2 \right)$ & $\left[ 3 \right]\,\left( 2 \right)$ & $\left[ 1 \right] \left( 1 \right)$ \\
\hline
\end{tabular}
\end{center}
The numbers in square brackets are the perturbative order of the various pieces (see sec.~[\ref{generalities}]) while the ones in round brackets are the numbers of bilinears in the term, with the convention that $1/2$ bilinear $=$ $1$ spinor.

Now, we have to give explicit algorithms to compute $C^{\left[ N \right]}$, $D_{1}^{\left[ N \right]}$, $D_{2}^{\left[ N \right]}$ and $D_{3}^{\left[ N \right]}$.

\subsubsection{$C^{\left[ N \right]}$}

Using the conventions given in sec.~[\ref{generalities}] we obtain the following result
\begin{align}
	\left( C^{\left[ N \right]}  \right)^{ R S}_{B'C'}
	= &
	\left[ e_{B'}^{ R}\, e_{C'}^{ S} \right]^{\left[ N \right]}
	=
	\nonumber\\
	= &
	\sum_{p=1}^{N}
	e_{B'}^{\left[ p \right] R}\, e_{C'}^{\left[ N-p+1 \right] S}
	\ .
	\label{algC}
\end{align}

\subsubsection{$D_{2}^{\left[ N \right]}$}

To obtain the $D_{2}^{\left[ N \right]}$ term we need $D_{1}^{\left[ N \right]}$ and the gauge field with flat index $A_{A}^{\left[ N \right]} = \left[ e_{A}^{ R} A_{ R} \right]^{\left[ N \right]}$.
For the former we have
\begin{align}
	\left( D_{1}^{\left[ N \right]}  \right)_{ R S\, A}
	= &
	- i \left[
		\bar\psi_{\left[  R \right.}
			\,\Gamma_{A}\,
		\psi_{\left.  S \right]}
	\right]^{\left[ N \right]}
	=
	\nonumber\\
	= &
	-i \sum_{p=1}^{N} 	
		\bar\psi^{\left[ p \right]}_{\left[  R \right.}
			\,\Gamma_{A}\,
		\psi_{\left.  S \right]}^{\left[ N-p+1 \right]}
		\ ,
	\label{algD1}
\end{align}
while the latter reads
\begin{align}
	A_{A}^{\left[ N \right]}
	= &
	\left[ e_{A}^{ R} A_{ R} \right]^{\left[ N \right]}
	= \nonumber \\
	= &
	\sum_{p=1}^{N} e_{A}^{\left[ p \right] R} A_{ R}^{\left[ N-p+1 \right]}
	\ .
	\label{AlgAflat}
\end{align}
Then, $D_{2}^{\left[ N \right]}$ becomes
\begin{align}
	\left( D_{2}^{\left[ N \right]}  \right)_{ R S}
	= &
	\left[ \left( D_{1} \right)_{ R S\, A}
			\,\eta^{AA'}\,A_{A'}  \right]^{\left[ N \right]}
	=
	\nonumber\\
	= &
	\sum_{p=1}^{N-1}\left( D_{1}^{\left[ p \right]} \right)_{ R S\, A}
	\,\eta^{AA'}\,A^{\left[ N-p \right]}_{A'}\ .
	\label{algD2}
\end{align}

\subsubsection{$ D_{3}^{\left[ N \right]} $}

Last, in analogy with (\ref{algD1}) we have
\begin{align}
	\left( D_{3}^{\left[ N \right]}  \right)_{ R S}
	= &
	 - i \left[
		\bar\psi_{\left[  R \right.}
		\psi_{\left.  S \right]}
	\right]^{\left[ N \right]}
	=
	\nonumber\\
	= &
	-i\sum_{p=1}^{N} 	
		\bar\psi^{\left[ p \right]}_{\left[  R \right.}
		\psi_{\left.  S \right]}^{\left[ N-p+1 \right]}
		\ .
	\label{algD3}
\end{align}

\subsection{Vielbein and Metric}

The vielbein is obtained as in eq.~(\ref{AlgSusyTransfv1})

\begin{align}
	\delta_{\epsilon} e_{M}^{\left[ N+1 \right]A}
		= &
		-\frac{1}{2} {\textrm{Re}}\left(  i \bar\epsilon \Gamma^{A} \psi_{M}^{\left[ N \right]} \right)
		\ ,
	\label{algVielbein}
\end{align}
then, the metric becomes
\begin{align}
	\delta_{\epsilon}^{\left[ N \right]} g_{M N}
	=  &
	\sum_{p=1}^{N} e^{\left[ p \right]\,A}_{\left( M \right.} \, e^{\left[ N-p+2 \right]\,B}_{\left.  N \right)} \eta_{AB}
	\ .
	\label{algMetric}
\end{align}

\subsection{Alternative Metric}

The metric is obtained from the susy transformation eq.~(\ref{AlgSusyTransfv1})
\begin{align}
	\delta_{\epsilon}^{\left[ N \right]} g_{M N}
	=  &
	- \frac{1}{2} {\textrm{Re}}\left[ i  \bar\epsilon \Gamma_{\left( M \right.}
	\psi_{\left. N \right)} \right]^{\left[ N \right]}
	= \nonumber\\ & =
	- \frac{1}{2} {\textrm{Re}}\left(  i  \sum_{p=1}^{N}  \bar\epsilon\, e_{\left( M \right.}^{\left[ p \right]A} \,\Gamma_{A}\,
	\psi^{\left[ N-p+1 \right]}_{\left. N \right)} \right)
	\ .
	\label{algMetricAlternative}
\end{align}

\subsection{Gauge Field}

Gauge field follows directly from eq.~(\ref{AlgSusyTransfv1})
\begin{equation}
	\delta^{\left[ N \right]}_{\epsilon}A_{M}
	=
	-\frac{\sqrt{6}}{4}{\textrm{Re}}
	\left( \bar\epsilon \psi^{\left[ N \right]}_{M} \right)
	\ .
	\label{algGauge}
\end{equation}

\section{Results}

In this section we collect the results obtained from the algorithms described in the previous section.\footnote{Notice that Fierz transformations (see appendix~[\ref{FIERZ}]) are used throughout the computation. }
First, we present the $AdS_{5}$ wigs constructed from one of the two Killing spinors $\eta_{0}$ and $\eta_{1}$.
Since each of them contains $4$ real degrees of freedom, the series truncates after the second order in bilinears.

The wig which depends only on $\eta_{0}$ turned out to be too simple: 
we show that it gives no contribution both to the  ADM mass and to the boundary stress--energy tensor.

$\eta_{1}\neq0$, $\eta_{0}=0$ case is more interesting: the explicit dependence on the boundary coordinates leads to a modification to BH Killing vectors. Furthermore, the boundary stress--energy tensor is not trivial and it will be discussed in section~[\ref{BSET}].

In order to present the result in different ways, we give the full wig and two particular limits of it, 
expanding in one case around small $\mu$ and in the other one around large $r$.
The former limit allows us to study a simplified, but a complete, metric while the latter shows the near boundary geometry.

The most general wig, obtained taking into account both $\eta_{0}$ and $\eta_{1}$, is derived. The degrees of freedom are now $8$, 
then the algorithm has to be iterated to the fourth order in bilinears.
The full expression is really cumbersome, even in the small $\mu$ and large $r$ limits. Then, we do not write it in this work, but the interested reader can find an electronic version in the ancillary files.

We repeat the procedure described above for the $AdS_{4}$ wigs. Apart from numerical coefficients, we find no substantial differences from the $AdS_{5}$ case. For this reason we present only the simplest results, leaving the complete wigs in the ancillary files.
Last remark, all wigs computed are asymptotically $AdS$.

\subsection{Results for $D=5$: $\eta_{1}=0$ and $\eta_{0}\neq 0$}

In this section we compute the finite BH wig choosing  $\eta_{1}=0$  and $\eta_{0}\neq 0$.
We introduce the following bilinears
\begin{align}
	\bM = -i \eta_{0}^{\dagger}\eta_{0}
	\ ,\qquad\qquad
	\bV_{i} = - i \eta_{0}^{\dagger}\hat\sigma_{i}\eta_{0}
	\ ,
	\qquad\qquad
	\lambda = \varepsilon_{0}^{t} \varepsilon_{0}
	\ ,
	\label{AlgBilEta0}
\end{align}
with these definitions, $\bM$ and $\bV_{i}$ are real numbers.

\subsubsection{Complete Wig}

The metric at first order is
\begin{align}
\delta^{\left[ 1 \right]}g
	= &\
-\frac{\mu}{r^2 h\left( r \right)} \lambda \bM \, \dd r\dd t
\ ,
	\label{AlgEta0ord1}
\end{align}
where we defined $h\left( r \right) = \sqrt{r^2+\frac{\mu}{r^{2}}}$.
The metric at second order is
\begin{align}
\delta^{\left[ 2 \right]}g
	= &\
-
\frac{1}{32 r^{4}}\left[ -3\mu^{2}+\mu r^{3}\left( -7 r+10h( r) \right)+12 r^{7}\left( r-h( r ) \right) \right]	\lambda^{2}\bM^2\, \dd t^2
+
\nonumber\\&
+
\frac{1}{32 r h(r)}\left[ \mu r\left( 14 r-3 h(r) \right)+16 r^{5}\left( r-h(r) \right) \right]	\lambda^{2}\bM^2\, \dd \vec{x}^2
+
\nonumber\\&
+
\frac{1}{32} \left( r h(r) \right)^{3/2} \left[ \mu r\left( 14 r-15 h(r) \right)+10 r^{5}\left( r-h(r) \right) \right]	\lambda^{2}\bM^2\,  \dd r^2
+
\nonumber\\&
-
\frac{1}{16 r^{2}} \left[\mu r \left( 3r+h(r) \right)+ 8r^{5}\left( r-h(r) \right) \right]	\lambda^{2}\bM \bV_{i}\, \dd t\dd x^{i}
\ .
	\label{AlgEta0ord2}
\end{align}
The gauge field is zero at every order.

\subsubsection{Expansion}

The complete metric result is now presented here in large-$r$ expansion and this coincides with the small-$\mu$ expansion.
\begin{align}
	\dd s^{2} = &\
	-\left( r^{2}+\frac{\mu}{r^{2}} \right)\, \dd t^{2}
	+\left( \frac{1}{r^{2}} - \frac{\mu}{r^{6}} \right) \dd r^{2}
	+r^{2} \dd \vec{x}^{2}
	-\frac{\mu}{r^2 h\left( r \right)} \lambda \bM \, \dd r\dd t	
	+
	\nonumber\\&\
-
	\frac{3\mu}{32}\lambda^{2} \bM^{2}\, \dd t^{2}
+
	\frac{3\mu}{32}\lambda^{2} \bM^{2}\, \dd\vec{x}^{2}
-
	\frac{3\mu}{16 r^{4}}\lambda^{2}\bM^{2}\, \dd r^{2}
-
\frac{3\mu^{2}}{32 r^{4}} \lambda^{2} \bM \bV_{i} \dd t \dd x^{i}
\ .
	\label{AlgEta0Expansion}
\end{align}

\subsubsection{ADM mass}

Following the procedure outlined in \cite{Behrndt:1998jd,Horowitz:1998ha}  we compute the ADM mass for the $\eta_{0}\neq0$, $\eta_{1}=0$ case.
The ADM mass is defined as
\begin{align}
	E_{ADM} = -\frac{1}{8\pi G} \int_{\Sigma} N \left( \Theta-\Theta_{0} \right)
	\ ,
	\label{Eta0ADM5dim2}
\end{align}
where $N=\sqrt{g_{tt}}$ is the norm of the timelike Killing vector $\partial_{t}$, $\Theta$ is the trace of the extrinsic curvature of a spacelike, near--infinity surface $\Sigma$ and $\Theta_{0}$ is $\Theta$ computed in the background $AdS_5$ geometry. Using the definition of extrinsic curvature we can rewrite eq.~(\ref{Eta0ADM5dim2}) as
\begin{align}
	E_{ADM} = -\frac{1}{8\pi G}  N \left( n^{\mu} - n^{\mu}_{0} \right)\partial_{\mu} A_{\Sigma}
	\ ,
	\label{Eta0ADM5dim21}
\end{align}
where $n^{\mu}$ is the vector normal to $\Sigma$ and $A_{\Sigma}$ is the area of $\Sigma$.
In order to consider a near infinity space--like surface, we use the large--$r$ metric eq.~(\ref{AlgEta0Expansion}).  We define a new radial coordinate
\begin{align}
	\rho^2 = r^{2} + \frac{3\mu}{32}\lambda^{2} \bM^{2}
	\ ,
	\label{Eta0ADM5dim1}
\end{align}
thus, the area of $\Sigma$ is simply $\rho^{3} V_{p}$, with $V_{p}$ the coordinate volume of the surface parametrized by $x^{i}$. The ADM mass is then
\begin{align}
	E_{ADM} = - \frac{3 \mu V_{p}}{16\pi G} + O\left( \frac{1}{\rho} \right)
	\ ,
	\label{Eta0ADM5dimFIN}
\end{align}
which is the result for Schwarzschild black hole. The wig constructed by bilinears only in $\eta_{0}$ gives no contribution to the ADM mass.

\subsubsection{Boundary Stress--Energy Tensor}

Using the prescription given in section~[\ref{BSET}] we compute the stress--energy tensor for the black hole wig. The result is
\begin{align}
	T_{\mu\nu} = -\frac{\mu}{2}\left( 4 u_{\mu}u_{\nu} +\eta_{\mu\nu} \right)
	\ ,
	\label{eta0LargeRTmunu}
\end{align}
where $u^{\mu}=\left( 1,0,0,0 \right)$ is the fluid velocity in the rest frame of the fluid. In this case, we have no contribution from the BH wig.

\subsection{Results for $D=5$: $\eta_{1}\neq0$ and $\eta_{0}= 0$}

In this section we compute the finite BH wig choosing  $\eta_{1}=0$  and $\eta_{0}\neq 0$. As in the previous case, we introduce
\begin{align}
	\bN = - i \eta_{0}^{\dagger}\eta_{0}
	\ ,\qquad\qquad
	\bK_{i} = - i \eta_{0}^{\dagger}\hat\sigma_{i}\eta_{0}
	\ ,
	\qquad\qquad
	\lambda = \varepsilon_{0}^{t} \varepsilon_{0}
	\ ,
	\label{AlgBilEta1}
\end{align}
where again $\bN$ and $\bK_{i}$ are real. Notice that in order to present the results we write the first terms in the large-$r$ expansion.

\subsubsection{First order in $\mu$}\label{d5mu}

As a first check, we want to determine only the effects due to gauge field and not to bilinears in the gravitino field. For this reason we consider the first order in the expansion around $\mu=0$ neglecting the contributions coming from bilinears in the gravitinos, since they contribute to  order $O\left( \mu^{2} \right)$.

The metric at first order is
\begin{align}
\delta^{\left[ 1 \right]}g
	= &\
	\, -\frac{\mu\lambda}{r^{2}}\left( \bN t+\bK_{i}x^{i} \right)\ \dd t^{2}
	+
	\frac{\mu\lambda}{r^{3}}\left[ -\bN\left( t^{2}+\vec{x}^{2} \right)-2tx_{i}\bK^{i} \right]\ \dd t \dd r
	+
	\nonumber\\ &
	-
	\frac{\mu\lambda}{2r^{2}}\left( t \bK_{i}+x_{i} N \right)\ \dd t \dd x^{i}
	+
	\frac{\mu\lambda}{2r^{5}}\bK_{i}\ \dd r \dd x^{i}
	-
	\frac{\mu\lambda}{2r^{2}}\left( \bN t+ \bK_{k}x^{k} \right)\delta_{ij}\ \dd x^{i} \dd x^{j}
\ .
	\label{AlgEta1ord1}
\end{align}
The metric at second order is
\begin{align}
\delta^{\left[ 2 \right]}g
	= &\
	-\frac{\mu}{2r^{4}} \lambda^{2} \bN \left[
	2 r^{2} t x^{i} \bK_{i}\left( 4+r^{2} \left( t^{2} + \vec{x}^{2} \right) \right)
	+ \bN \left( 1+r^{2} \left( t^{2} -3  \vec{x}^{2} \right) \right)
	\right]	\, \dd t^2
+
\nonumber\\&
-\frac{2\mu}{r^{6}}\lambda^{2} \bN \left[
\bN t^{2} + r^{2} t x^{i} \bK_{i} \left( t^{2}+\vec{x}^{2} \right)
\right]\, \dd r^2
+
\nonumber\\&
-\frac{\mu}{r^{5}} \lambda^{2} \bN \left[
t \bN \left( 2+ r^{2} \left( t^{2} -  \vec{x}^{2} \right) \right)
+ x^{i}\bK_{i} \left( 1 + 2 r^{2} \left(  3 t^{2} +\vec{x}^{2} \right) \right)
\right]
\, \dd t\dd r
+
\nonumber\\&
+\frac{\mu}{4 r^{4}}\lambda^{2} \bN
\left[
-2 r^{2} x_{i} \left( 3 t \bN + 2 x^{j} \bK_{j} \right)
+
\bK_{i} \left( 1+ r^{2}\left( -3t^{2}+\vec{x}^{2} \right) \right)
\right]
\, \dd t\dd x^{i}
+
\nonumber\\&
-\frac{\mu}{2 r^{5}} \lambda^{2} \bN
\left[
x_{i}\left( \bN+8 r^{2} t x^{j} \bK_{j} \right)
+
t \bK_{i} \left( -1+2r^{2}\left( t^{2} -\vec{x}^{2} \right) \right)
\right]
\, \dd r\dd x^{i}
+
\nonumber\\&
+\frac{\mu}{4 r^{5}}\lambda^{2} \bN
\left[
\bN r \left( -1 +r^{2} \left( t^{2} +3 \vec{x}^{2} \right) \right) \delta_{i j}
+
4 r^{3} t x^{k} \bK_{k} \left( -1 + r^{2} \left( t^{2}+\vec{x}^{2} \right) \right)\delta_{i j} +
\right.
\nonumber\\&
\left.\qquad\qquad\quad
-2 \bN r^{3} x_{i} x_{j} + r^{3} t x_{i} \bK_{j}
\right]
\, \dd x^{i} \dd x^{j}
	\ .
	\label{AlgEta1ord2}
\end{align}
In this limit, the gauge field is zero at each order.

\subsubsection{Large $r$ expansion}

Here we compute the large-$r$ expansion of the metric corrections.

At first order, we have
\begin{align}
\delta^{\left[ 1 \right]}g
	= &\
	-\frac{\mu}{r^{2}} \lambda
	\left[
	t \bN + x^{i} \bK_{i}
	\right]
		\ \dd t^{2}
	-
	\frac{\mu}{r^{3}} \lambda
	\left[
	2 t x^{i} \bK_{i} + \bN \left( t^{2} + \vec{x}^{2} \right)
	\right]
		\ \dd t \dd r
	+
	\nonumber\\ &
	-
	\frac{\mu}{2 r^{2}} \lambda\left( t \bK_{i} +x_{i} \bN  \right)
		\ \dd t \dd x^{i}
	+
	\frac{\mu}{2 r^{5}} \bK_{i}
		\ \dd r \dd x^{i}
	-
	\frac{\mu}{2 r^{2}} \lambda \left( t \bN + x^{k} \bK_{k} \right) \delta_{ij}
	\ \dd x^{i} \dd x^{j}
\ .
	\label{AlgEta1LargeR}
\end{align}
The metric at second order is
\begin{align}
\delta^{\left[ 2 \right]}g
	= &\
	-\mu \lambda^{2} \bN t x^{i} \bK_{i} \left( t^{2} +\vec{x}^{2} \right)
	\, \dd t^2
+
\nonumber\\&
-\frac{2\mu}{r^{4}} \lambda^{2} t x^{i} \bN \bK_{i} \left( t^{2} +\vec{x}^{2} \right)
	\, \dd r^{2}
+
\nonumber\\&
-\frac{\mu}{ r^{3}} \lambda^{2} \bN
\left[
t \bN \left( \left( t^{2} - \vec{x}^{2} \right) \right) +2 x^{i} \bK_{i} \left( 3 t^{2} +\vec{x}^{2} \right)
\right]
	\, \dd t\dd r
+
\nonumber\\&
+\frac{\mu}{4 r^{2}} \lambda^{2} \bN
\left[
-2 x_{i} \left( 3 \bN t + 2 x^{k} \bK_{k} \right) +
\bK_{i} \left( \left(- 3  t^{2} +\vec{x}^{2} \right) \right)
\right]
	\, \dd t\dd x^{i}
+
\nonumber\\&
-\frac{\mu}{r^{3}} \lambda^{2} \bN
\left[
4 t^{2} x_{i} x^{k} k_{k}
+
\bK_{i} \left( t^{2} - \vec{x}^{2} \right)
\right]
	\, \dd r\dd x^{i}
+
\nonumber\\&
+
\mu \lambda^{2} \bN
\left[
t x^{k }\bK_{k} \left( t^{2} +\vec{x}^{2} \right)\delta_{ij}
+
\frac{1}{2r^{2}} \left(
t x_{i} \bK_{j} -2 \bN x_{i} x_{j}
\right)
\right]
	\, \dd x^{i} \dd x^{j}
	\ .
	\label{AlgEta1ord2LargeR}
\end{align}
The only non--zero components of the gauge field are the $A^{\left[ 2 \right]}_{i}$
\begin{align}
	A^{\left[ 2 \right]}_{i}
	=&\
	\frac{3\sqrt{6}\mu^{2}}{256 r^{6}}\lambda^{2}
	\varepsilon_{ijk} x^{j} \bN \bK^{k}\left( t^{2}+\vec{x}^{2} \right) \ .
	\label{gaugeFieldlargeR4}
\end{align}

\subsubsection{Complete}

Here we present the complete wig depending on $\eta_{1}$ bilinears. The first order is
\begin{align}
\delta^{\left[ 1 \right]}g
	= &\
	-
	\frac{\mu}{r^{3}} \lambda h\left( r \right)
	\left[
	t \bN + x^i \bK_i	
	\right]
	\ \dd t^{2}
	-
	\frac{\mu}{r^{2}h\left( r \right)} \lambda
	\left[
		\left( t^{2}-\vec{x}^{2} \right) \bN
		+2 t x^{i} \bK_{i}	
		\right]
	\ \dd t \dd r
	+
	\nonumber\\ &
	+
	\lambda r \left( r- h\left( r \right) \right)
	\left[ t \bK_{i}+x_{i} \bN \right]
	\ \dd t \dd x^{i}
	-
	\frac{1}{r h\left( r \right)}\left( r-h\left( r \right) \right) \bK_{i}
	\ \dd r \dd x^{i}
	+
	\nonumber\\ &
	+
	\lambda r \left( r-h\left( r \right) \right)
	\left[
	t \bN + x^i \bK_i	
	\right] \delta_{ij}
	\ \dd x^{i} \dd x^{j}
\ .
	\label{AlgEta1ord1full}
\end{align}
The second order is
{\allowdisplaybreaks
\begin{align}
\delta^{\left[ 2 \right]}g
	= &\
	-\frac{1}{16 r^{8}}
	\bN\lambda^2
	\left[
		2 r^2 tx^{i} \bK_{i}\left(-3\mu ^2\left(-11+4 r^2\left( t^2+\vec{x}^{2}\right)\right)+3\mu  r^4\left(9+4 r^2\left( t^2+\vec{x}^{2}\right)\right)
		+
		\right.\right.\nonumber\\& \left.\left.
		+2\mu  r^3 h\left(r\right)+2 r^6\left(-3+4 r^2\left( t^2+\vec{x}^{2}\right)\right)\left( r^2- r h\left(r\right)\right)\right)
		+
		\right.\nonumber\\& \left.+
		\bN\left(
		\mu ^2\left(
			11+6 r^4\left(- t^2+\vec{x}^{2}\right)^2
		+
		r^2\left(13 t^2-31\vec{x}^{2}\right)
			\right)
		+
		\right.\right.\nonumber\\& \left.\left.
		+2 r^6\left(
			-3+6 r^4\left(- t^2+\vec{x}^{2}\right)^2+ r^2\left(-25 t^2+11\vec{x}^{2}\right)
			\right)\left(- r^2+ r h\left(r\right)\right)
	+\right.\right.\nonumber\\& \left.\left.
		+\mu \left(
			2 r^8\left(- t^2+\vec{x}^{2}\right)^2
			-6 r^3 h\left(r\right)
			+
			r^4\left(17+2 r\left( t^2+\vec{x}^{2}\right) h\left(r\right)\right)
		+\right.\right.\right.\nonumber\\& \left.\left.\left.
					- r^6\left(
				8 r t^4 h\left(r\right)+\vec{x}^{2}\left(37+8 r\vec{x}^{2} h\left(r\right)\right)
				-
				t^2\left(31+16 r\vec{x}^{2} h\left(r\right)\right)
				\right)
			\right)
		\right)
	\right]
\ \dd t^{2}
	+
	\nonumber\\ &
	-\frac{1}{16 r^{6} h\left( r \right)}
\bN\lambda^2\left[
		\bN t\left(
		2 r^4\left(31+ r^2\left( t^2-\vec{x}^{2}\right)\right)\left(- r^2+ r h\left(r\right)\right)
			+
		\right.\right.\nonumber\\& \left.\left.
	+	\mu \left(
	22 r h\left(r\right)+ r^2\left(-21+5\left( t^2-\vec{x}^{2}\right)\left(- r^2+4 r h\left(r\right)\right)\right)
				\right)
			\right)
			+
		\right.\nonumber\\& \left.
		+x^{i} \bK_{i}\left(
		10 r^4\left(-1+ r^2\left(3 t^2+\vec{x}^{2}\right)\right)\left(- r^2+ r h\left(r\right)\right)
			+\mu\left(
			14 r h\left(r\right)
				+
		\right.\right.\right.\nonumber\\& \left.\left.\left.
			+ r^2\left(7+9\left(3 t^2+\vec{x}^{2}\right)\left(- r^2+4 r h\left(r\right)\right)\right)
			\right)
		\right)
	\right]
\ \dd t \dd r
	+
	\nonumber\\ &
	+
	\frac{1}{8 r^7\left( h\left(r\right)\right)^{3}}
\bN\lambda^2\left[
		4 r^2 tx^{i} \bK_{i}\left(2 r^6-2 r^5 h\left(r\right)
		+\right.\right.\nonumber\\& \left.\left.
		+\mu \left(- r h\left(r\right)+ r^2\left(2-\left( t^2+\vec{x}^{2}\right)\left(- r^2+5 r h\left(r\right)\right)\right)\right)\right)
		+
		\right.\nonumber\\& \left.	+
		\bN\left(
			2 r^4\left(5+2 r^2\left(- t^2+\vec{x}^{2}\right)+2 r^4\left(- t^2+\vec{x}^{2}\right)^2\right)\left( r^2- r h\left(r\right)\right)
			+\right.\right.\nonumber\\& \left.\left.
			+\mu \left(
				5 r^6\left(- t^2+\vec{x}^{2}\right)^2-5 r h\left(r\right)+2 r^2\left(5+ r\left(-5 t^2+\vec{x}^{2}\right) h\left(r\right)\right)
			+\right.\right.\right.\nonumber\\& \left.\left.\left.	- r^4\left(3 r t^4 h\left(r\right)+3 r\vec{x}^{4} h\left(r\right)-2 t^2\left(-4+3 r\vec{x}^{2} h\left(r\right)\right)\right)
			\right)
		\right)
	\right]
	\ \dd r^{2}
	+
	\nonumber\\ &
	+
	\frac{1}{32 r^{5}}
\bN\lambda^2\left[
	\bK_{1}\left(\mu  r\left(6+ r^2\left(3 t^2+\vec{x}^{2}\right)+3 r^4\left(- t^4+\vec{x}^{4}\right)\right)
	+\right.\right.\nonumber\\& \left.\left.
	+2 r^5\left(3+ r^2\left(9 t^2-5\vec{x}^{2}\right)+4 r^4\left(- t^4+\vec{x}^{4}\right)\right)\right)
	+\right.\nonumber\\& \left.
		-2 r^3x_{1}\left(
			\bN\left(13\mu -2 r^4\right) t+ x^{i} \bK_{i}\left(-14 r^4+8 r^6\left(3 t^2+\vec{x}^{2}\right)
		+\right.\right.\right.\nonumber\\& \left.\left.\left.	
			+\mu \left(-1+3 r^2\left(3 t^2+\vec{x}^{2}\right)\right)\right)\right)
			-\left(
		\bK_{1}\left(\mu \left(-1+2 r^2\left(3 t^2+\vec{x}^{2}\right)
		+\right.\right.\right.\right.\nonumber\\& \left.\left.\left.\left.
		+ r^4\left( t^4-\vec{x}^{4}\right)\right)+2 r^4\left(3+ r^2\left(9 t^2-5\vec{x}^{2}\right)+4 r^4\left(- t^4+\vec{x}^{4}\right)\right)\right)
		+\right.\right.\nonumber\\& \left.\left.	
		+
			2 r^2x_{1}\left(
				2\bN\left(-\mu + r^4\right) t+ x^{i} \bK_{i}\left(14 r^4-8 r^6\left(3 t^2+\vec{x}^{2}\right)
				+\right.\right.\right.\right.\nonumber\\& \left.\left.\left.\left.
				+\mu \left(2+ r^2\left(3 t^2+\vec{x}^{2}\right)\right)\right)
			\right)
		\right) h\left(r\right)
	\right]
	\ \dd t \dd x^{i}
	+
	\nonumber\\ &
	+
	\frac{1}{32 r^{6}\left( h\left(r\right)\right)^3}
\bN\lambda^2\left[
	-2\left(\mu + r^4\right)\left(
				-x_{1}\left(2 x^{i} \bK_{i} r^2\left(-15\mu +7 r^4\right) t
			+\right.\right.\right.\nonumber\\& \left.\left.\left.+					
					\bN\left(11 r^4+25 r^6\left(- t^2+\vec{x}^{2}\right)+\mu \left(2+9 r^2\left(- t^2+\vec{x}^{2}\right)\right)\right)
					\right)
			+\right.\right.\nonumber\\& \left.\left.	+
				\bK_{1} t\left(13 r^4+\mu \left(4+ r^2\left(7 t^2-9\vec{x}^{2}\right)\right)- r^6\left(5 t^2+\vec{x}^{2}\right)\right)\right)
		+
		\right.\nonumber\\& \left.	
		- r^3\left(
			\bK_{1} t\left(-26 r^4+2 r^6\left(5 t^2+\vec{x}^{2}\right)+\mu \left(-29+ r^2\left(7 t^2+3\vec{x}^{2}\right)\right)\right)
		+\right.\right.\nonumber\\& \left.\left.	+
			x_{1}\left(2 x^{i} \bK_{i} r^2\left(9\mu +14 r^4\right) t
		+\right.\right.\right.\nonumber\\& \left.\left.\left.	+
				\bN\left(22 r^4+50 r^6\left(- t^2+\vec{x}^{2}\right)+\mu \left(23+43 r^2\left(- t^2+\vec{x}^{2}\right)\right)\right)
			\right)
		\right) h\left(r\right)
	\right]
	\ \dd r \dd x^{i}
	+
	\nonumber\\ &
	+\frac{1}{16 r^6 h\left( r \right)^{2}}
	\bN\lambda^2\left[
	-\left(\mu + r^4\right)\left(-4 x^{k} \bK_{k} r^2 t\left(13 r^4+\mu \left(1+5 r^2\left( t^2+\vec{x}^{2}\right)\right)\right)
		+\right.\right.\nonumber\\& \left.\left.	
	+
	\bN\left(4\left( r^4-4 r^6\left( t^2-\vec{x}^{2}\right)+3 r^8\left( t^2-\vec{x}^{2}\right)^2\right)
	+\right.\right.\right.\nonumber\\& \left.\left.\left.
	+
	\mu \left(6+5 r^4\left( t^2-\vec{x}^{2}\right)^2+ r^2\left(-11 t^2+3\vec{x}^{2}\right)\right)\right)\right)
	+\right.\nonumber\\& \left.
	+
	r^3\left(-2 x^{k} \bK_{k} r^2 t\left(26 r^4+\mu \left(23+2 r^2\left( t^2+\vec{x}^{2}\right)\right)\right)
	+\right.\right.\nonumber\\& \left.\left.
	+
	\bN\left(4\left( r^4-4 r^6\left( t^2-\vec{x}^{2}\right)+3 r^8\left( t^2-\vec{x}^{2}\right)^2\right)
	+\right.\right.\right.\nonumber\\& \left.\left.\left.
	+
	\mu \left(4+11 r^4\left( t^2-\vec{x}^{2}\right)^2+ r^2\left(-15 t^2+23\vec{x}^{2}\right)\right)\right)\right) h\left(r\right)
	\delta_{ij}
	+
	\right.\nonumber\\& \left.
-\left(\mu + r^4\right)\left(+\left(-\bN\mu +8 x^{k} \bK_{k} r^2\left(\mu +2 r^4\right) t\right)x_{i}x_{j}
+\right.\right.\nonumber\\& \left.\left.
+
\left(\bK_{i}x_{j}+\bK_{j}x_{i}\right) t\left(\mu \left(-3+4 r^2\left( t^2-\vec{x}^{2}\right)\right)
+
2 r^4\left(7+4 r^2\left( t^2-\vec{x}^{2}\right)\right)\right)\right)
+\right.\nonumber\\& \left.
+
 r^3\left(\left(-5\bN\mu +16 x^{k} \bK_{k} r^2\left(\mu + r^4\right) t\right)x_{i}x_{j}
 +\right.\right.\nonumber\\& \left.\left.
 +
 2\left(\bK_{i}x_{j}+\bK_{j}x_{i}\right) t\left(4\mu \left(1+ r^2\left( t^2-\vec{x}^{2}\right)\right)
 +\right.\right.\right.\nonumber\\& \left.\left.\left.
 +
 r^4\left(7+4 r^2\left( t^2-\vec{x}^{2}\right)\right)\right)\right) h\left(r\right)\right]
	\ \dd x^{i} \dd x^{j}
\ .
	\label{AlgEta1ord2full}
\end{align}
}

\subsection{Results for $D=4$: $\eta_{1}=0$ and $\eta_{0}\neq 0$}

The $AdS_{4}$ model is very similar to $AdS_{5}$ one. For our purpose, the only relevant difference is the Schwarzschild BH metric
\begin{align}
	\dd s^{2} = f(r)^2 \dd t^{2} + f(r)^{-2} \dd r^{2} + r^{2} \dd \vec{x}^{2}\ ,
	\label{AdSmetric}
\end{align}
where $f(r)=\sqrt{r^{2}+\frac{\mu}{r}}$.
Due to the fact that $2$-- and $3$--dimensions spinors have the same number of degrees of freedom, our algorithm can be applied with no modifications. Notice also that the Killing spinors are written in the same way of eq.~(\ref{KSsolution2}), where $x^{i}$ denotes only $x_{1}$ and $x_{2}$.

Last remark, in $4d$  $\Gamma_{5}$  is defined by dimensional reduction from $5d$ as
\begin{align}
	\Gamma_{5} = &\ \sigma_{3}\otimes\hat\sigma_{3}
	\ ,
	\label{gamma5in4d}
\end{align}
then, bilinears in $\eta$ with $\hat\sigma_{3}$ are still present.

\subsubsection{Complete Wig}

The first order is
\begin{align}
\delta^{\left[ 1 \right]}g
=&\
-\frac{3\mu}{4 r f\left( r \right)} \lambda \bM
\ \dd r \dd t
\ .
\end{align}
The second order is
\begin{align}
\delta^{\left[ 2 \right]}g
=&\
\frac{1}{32 r^{2}} \lambda^{2} \bM^{2}
\left( 7r^{3}-2\mu \right)
\left[ \mu + 2 r^{2} \left( r-f\left( r \right) \right)\right]\
\dd t^2
+\nonumber \\
&
- \frac{1}{64 f}\lambda^{2} \bM^{2}
\left[
11\mu^{2} + 14 \mu r^{3}
+r^{2}\left( 25\mu+28r^{3} \right)\left( r-f\left( r \right) \right)
\right]
\delta_{ij} \dd x^i \dd x^j
+ \nonumber \\
&
-
\frac{3}{64}\lambda^{2} \bM \bV_{i}
\left[
4 \mu r + \left( \mu + 8 r^{3} \right)\left( r - f\left( r \right) \right)
\right]
\dd t \dd x^i +
\nonumber\\
&
+
\frac{3}{32 r f\left( r \right)^{3}} \lambda^{2} \bN^{2}
\left[
2 \mu r + \left( 3\mu +4 r^{3}\right)\left( r-f\left( r \right) \right)
\right]\
\dd r^2
\ .
\end{align}
Notice that, as in the $5$--dimensional case, the gauge field is zero at every order.

\subsection{Results for $D=4$: $\eta_{1}\neq0$ and $\eta_{0}= 0$}

In this section we compute the finite wig choosing  $\eta_{1}=0$  and $\eta_{0}\neq 0$.
We introduce the following bilinears
\begin{align}
	\bN =-i \eta_{0}^{\dagger}\eta_{0}
	\ ,\qquad\qquad
	\bK_{i} = -i \eta_{0}^{\dagger}\hat\sigma_{i}\eta_{0}
	\ ,
	\qquad\qquad
	\lambda = \varepsilon_{0}^{t} \varepsilon_{0}
	\ ,
	\label{AlgBilEta11}
\end{align}
with these definitions, $\bN$ and $\bK_{i}$ are real quantities.

\subsubsection{First order in $\mu$}

As in [\ref{d5mu}], we focus on effects due to gauge field and not to bilinears in the gravitino field, considering only the first order in the expansion around $\mu=0$.
The metric at first order is
\begin{eqnarray}
\delta^{\left[ 1 \right]}g
= &-&\frac{\l \m}{2r^2} \left(\bN t + \bK_i x^i \right) \dd t^2 -\frac{\l \m}{4r^4}\left[6r^2 t \left( \bK_i x^i\right)+ \bN\left(-1+3r^2\left(t^2+\vec{x}^{2}\right)\right)\right] \dd t \dd r \nonumber \\
&-&\frac{\l \m}{2r}\left(\bK_i t + \bN x_i\right)\dd t \dd x^i+ \frac{\l \m}{2r^4} \bK_i \dd r \dd x^i -\frac{\l \m}{2r} \left(\bN t +\bK_k x^k\right) \delta_{ij} \dd x^i \dd x^j
\ .
\nonumber\\
	\label{AlgEta1ord14dim}
\end{eqnarray}
The metric at second order is
\begin{eqnarray}
\delta^{\left[ 2 \right]}g
	= &-&\frac{\m}{8r^3}
	\lambda^{2} \bN
	\left[2r^2 t \left(\bK_i x^i\right)\left[7+3r^2\left(t^2 + \vec{x}^{2}\right)\right]+\bN\left[1+r^2\left(t^2-5 \vec{x}^{2}\right)\right]\right] \dd t^2
+
\nonumber \\
&-& \frac{3 \m t}{4r^5} \lambda^{2} \bN \, t
\left[\bN t +\left(\bK_k x^k\right)\left[-1 + r^2\left(t^2 + \vec{x}^{2}\right)\right]\right] \dd r^2
+
\nonumber \\
&-& \frac{\m}{8r^4}\lambda^{2} \bN
\left[\bN t \left[5+3r^2\left(t^2-x_i x^i\right)\right]+2\left(\bK_i x^i\right)\left[1+3r^2\left(3t^2+\vec{x}^{2}\right)\right]\right] \dd t \dd r
+
\nonumber \\
&-& \frac{\m}{4r} \lambda^{2} \bN
\left[ x_i \left(3 \bN t + 2 \bK_k x^k\right) + 2 \bK_i t^2\right] \dd t \dd x^i
+
\nonumber \\
&-&\frac{\m}{4r^4} \lambda^{2} \bN
\left[ \left( \bN x_i - t \bK_i \right) + r^2 t \left[  3 \bK_i  \left(   t^2-\vec{x}^2 \right)   +  12 x_i x^j \bK_j \right]\right] \dd r \dd x^i
+
\nonumber \\
&-&
\frac{\mu}{8 r^{3}} \lambda^{2} \bN \left[
	\left( -1 + r^{2}\left( t^{2}+5\vec{x}^{2} \right) \right)\delta_{ij}
	+6 r^{2} t x^{k} \bK_{k} \left[ -1+r^{2}\left( t^{2}+\vec{x}^{2} \right) \right]\delta_{ij}
	+
	\right.
	\nonumber\\&&
	\left.\qquad\quad
	-4 r^{2} \bN x_{i}x_{j}
	+2 r^{2} t x_{i} \bK_{j}
\right]
\ \dd x^{i} \dd x^{j}
	\ .
	\label{AlgEta1ord24dim}
\end{eqnarray}
For both orders, the gauge field is zero.

\subsubsection{Large-$r$ expansion}

Here we compute the large-$r$ expansion of the metric corrections.

The first order metric is
\begin{eqnarray}
\delta^{\left[ 1 \right]}g
= &-&\frac{\l \m}{2r}\left(\bN t + \bK_i x^i\right) \dd t^2 -\frac{3\l \m}{4r^2}\left[2t \left(\bK_i x^i \right) +\bN\left(t^2 + \vec{x}^{2}\right)\right] \dd t \dd r
	+ \nonumber \\ &-&
	\frac{\l \m}{2r}\left(\bK_i t + \bN x_i\right) \dd x^i \dd t
	-
	\frac{\l \m}{2r^4} \bK_i \dd x^i \dd r
	+ \nonumber \\ &-&
	\frac{\l \m}{2r}\left(\bN t +\bK_i x^i \right) \delta_{ij}\dd x^i \dd x^j
\ .
	\label{AlgEta1LargeR4dim}
\end{eqnarray}
The second order is
\begin{eqnarray}
\delta^{\left[ 2 \right]}g
	= &-&\frac{3\mu}{4}\l^2  \bN r t\left(\bK_i x^i\right)\left(t^2+\vec{x}^{2}\right) \dd t^2
	-
	\frac{3 \mu}{4r^3} \l^2 \bN t \left(\bK_i x^i\right)\left(t^2 +\vec{x}^{2}\right) \dd r^2
	+ \nonumber \\
&-&\frac{3  \m}{8 r^2}\l^2 \bN \left[\bN t\left(t-\vec{x}^{2}\right)+2\left(\bK_i x^i\right)\left(3t^2+\vec{x}^{2}\right)\right] \dd t \dd r
+ \nonumber \\
&-&\frac{\mu}{4r}\l^2 \bN\left[2\bK_i t^2 + x_i\left(x^k \bK_k  + 3 t \bN\right)\right] \dd t \dd x^i
+ \nonumber \\ &-&
\frac{3 \mu}{4r^2} \l^2 \bN t \left[4x_i\left(\bK_k x^k\right)-\bK_i\left(t^2 +\vec{x}^{2}\right)\right] \dd r \dd x^i
+ \nonumber \\
&+& \frac{\mu}{4} \l^2 \bN \left[3r t \left(\bK_k x^k\right)\left(t^2+\vec{x}^{2}\right) \delta_{ij} + \frac{2}{r}\left(t \bK_i x_j - 2 \bN x_i x_j\right)\right]\dd x^i \dd x^j
	\ .
	\label{AlgEta1ord2LargeR4dim}
\end{eqnarray}
The  non--zero components of the guge field are the $A^{\left[ 2 \right]}_{i}$
\begin{align}
	A^{\left[ 2 \right]}_{i}
	=&\
	\frac{\sqrt{6}\mu^{2}}{128 r^{4}}\lambda^{2}
	\varepsilon_{ij} x^{j} \bN \bK_{3}\left( -t^{2}+\vec{x}^{2} \right)
	\ ,
	\label{gaugeFieldlargeR5}
\end{align}
where $\varepsilon_{ij}$ is the $2d$ antisimmetric tensor, with $\varepsilon_{12}=1$.

\section{Boundary Stress--Energy Tensor}\label{BSET}

We now proceed calculating the stress--energy tensor dual to the black hole using the prescription given in \cite{Brown:1992br}, \cite{Balasubramanian:1999re}. First of all we define the constrain that will allow to foliate the spacetime in slices at constant $r$
\begin{equation}
\Phi = r - c = 0	
\ ,
	\label{BYconstraint}
\end{equation}
with $c \in \mathbb{R}$.\footnote{We adopt the notation given in \cite{Poisson}} The outward-pointing normal vector to the boundary $\left.\mathcal{M}\right|_{r = c}$ is defined as\footnote{This definition is valid as long as the surface is not null-like. In that case, the outward pointing normal will be $k_{M} = - \partial_{M} \Phi$. See \cite{Poisson} for further details.}
\begin{equation}
n_{M} = \frac{\partial_{M}\Phi}{\sqrt{g^{R S}\partial_{R}\Phi\partial_{S}\Phi}}	
\ .
	\label{BYnormal}
\end{equation}
Using $n_{M}$ we define the boundary metric $\gamma$:
\begin{equation}
\hat{\gamma}_{M N} = g_{M N} - n_{M} n_{N}
\ .
	\label{BYboundarymetric}
\end{equation}
In order to obtain a $4-$dimensional metric we have to eliminate from $\hat{\gamma}$ the first column and the first row 
\begin{equation}
\gamma_{MN} = \left( \begin{array}{c|cc}
\gamma_{rr} & \gamma_{rt} & \gamma_{rj} \\
\hline
\gamma_{tr} & \multicolumn{2}{c}{\multirow{2}{*}{$\gamma_{\mu\nu}$}}  \\
\gamma_{ir} &  &
\end{array} \right)	\ .
	\label{BYfourdimmetric}
\end{equation}
In a similar fashion we calculate the extrinsic curvature $\Theta_{MN}$ and then $\Theta_{\mu\nu}$:
\begin{equation}
\Theta_{M N} = -\frac{1}{2} \left(\nabla_M n_N + \nabla_N n_M \right) \ .
	\label{BYextrinsicCurv}
\end{equation}
Finally we can define our (boundary) stress energy tensor as
\begin{equation}
	T^{\mu\nu} = \frac{1}{8 \pi G} \left(\Theta^{\mu\nu} - \Theta \gamma^{\mu\nu} - 3 \gamma^{\mu\nu} - \frac{1}{2}G^{\mu\nu}\right) \ ,
	\label{BYTmunu}
\end{equation}
where $\Theta$ is defined as the trace of $\Theta^{\mu\nu}$ and $G^{\mu\nu}$ is the Einstein tensor\footnote{A careful reader may have noticed a change of sign in front of the Einstein tensor with respect to \cite{Balasubramanian:1999re}. This is just a matter of convention in the definition of the Riemann tensor.} build from $\gamma^{\mu\nu}$. Note that we set $R_{AdS} = 1$  as usual.

\subsection{Stress--Energy Tensor for $AdS_{5}$}

Using the prescription given in the previous section we present the result obtained in $AdS_5$. The first-order corrections are the same both at 4 and 5 dimensions, while at second-order one they are different. We decompose the contribution to the stress-energy tensor in the perturbative form as 
 \begin{align}
	T_{\mu\nu} = -\frac{\mu}{2}\left( 4 u_{\mu}u_{\nu} +\eta_{\mu\nu} \right) + \lambda\mu\mathcal{T}^{\left[ 1 \right]}_{\mu\nu} + \lambda^{2}\mu\mathcal{T}^{\left[ 2 \right]}_{\mu\nu}
	\ ,
	\label{eta1LargeRTmunu5dim}
\end{align}
where $u^{\mu}=\left( 1,0,0,0 \right)$ is the fluid velocity in the rest frame of the fluid as usual, $\bB^{\m}=\left(-\bN,\bK_i\right)$ is the bilinear $4-$vector. As usual, 
we define the projectors
\begin{align}
	& P^{\shortparallel}_{\mu\nu}= \eta_{\mu\nu} + u_{\mu} u_{\nu}
	 \ ,
	 &P^{\bot}_{\mu\nu}= - u_{\mu} u_{\nu}
	 \ .
	\label{EMT}
\end{align}
The first order of $T_{\m\nu}$ is
\begin{eqnarray}
	\mathcal{T}^{\left[ 1 \right]}_{\mu\nu} &=& -\frac{d}{4}  \left[(\bB \cdot x)( \eta_{\mu\nu}+ d\,  u_\mu u_\nu )  -  4 P^{\shortparallel}_{\left(  \mu \right. \rho}  P^{\bot}_{\left. \nu \right) \sigma} \bB^{\left[  \rho  \right.} x^{\left. \sigma \right]}\right] \ ,
	\label{eta1BY2}
\end{eqnarray}
where $d$ refers to $AdS_{d+1}$. Notice that the second term in eq.~(\ref{eta1BY2}) resembles a vorticity term. Actually, the relativistic vorticity term is defined as
\begin{equation}
\Delta_{\mu\nu} = P^{\shortparallel}_{\mu\lambda}P^{\shortparallel}_{\nu\tau} \nabla^{\left[\lambda\right.}u^{\left.\tau\right]} \ .
\end{equation}
In our case the second spatial projector is actually an orthogonal projector, that in fact, mixes space and time components as a result of the supersymmetry. 
$\bB$ may be seen as a ``super-correction'' to fluid velocity. However, a deeper analysis is due.

The second order reads
\begin{eqnarray}	
	\mathcal{T}^{\left[ 2 \right]}_{\mu\nu} &=& P^{\bot}_{\mu\nu}\left[\left(\frac{5}{2}+y\right)\left(\bB\cdot x\right)^2 + 12\left(x\bB\right)^{\bot}\left(x\bB\right)^{\shortparallel}+\left(x\bB\right)^{\bot}\left(xx\right)^{\shortparallel}\right] + \nonumber \\
&-&P^{\shortparallel}_{\mu\nu}\left[\left(\frac{1}{2}+y\right)\left(\bB\cdot x\right)^2+\frac{9}{2}\left(x\bB\right)^{\bot}\left(x\bB\right)^{\shortparallel}+\left(\bB \bB\right)^{\bot}\left(xx\right)^{\shortparallel}\right]+ \nonumber \\
&+&y \left(\bB\cdot x\right)^2\left(\eta_{\mu\nu}+4u_{\mu}u_{\nu}\right)
+
\left(\bB \bB\right)^{\bot}x_{\left(\mu\right.}x_{\left.\nu\right)}-2\left(x\bB\right)^{\bot}\bB_{\left(\mu\right.}x_{\left.\nu\right)} + \nonumber \\
&-&P^{\shortparallel}_{\left(\mu\rho\right.}P^{\bot}_{\left.\nu\right)\sigma}\left\{-4\bB^{\left[\rho\right.}x^{\left.\sigma\right]}\left[4 \bB \cdot x +\frac{3}{2}\left(x\bB\right)^{\bot} \right]
-x^{\rho}x^{\sigma}\left(\bB \bB\right)^{\bot}+\bB^{\rho}\bB^{\sigma}\left(xx\right)^{\shortparallel}\right\}
\ , \nonumber \\
	\label{eta1BY2}
\end{eqnarray}
with
\begin{equation}
(VW)^{\bot}= P^{\bot}_{\mu\nu}V^{\mu}W^{\nu} \ , \qquad \qquad (VW)^{\shortparallel}= P^{\shortparallel}_{\mu\nu}V^{\mu}W^{\nu} \ ,
\end{equation}
and we have used Fierz identities (see appendix~[\ref{FIERZ}]) to substitute
\begin{equation}
\left(\bB\cdot x\right)^2 = \left[2\left(x\bB\right)^{\bot}\left(x\bB\right)^{\shortparallel} + x^2 \left(\bB \bB\right)^{\bot}\right] \ .
\end{equation}
The coefficient $y$ can be set in order to recast $\mathcal{T}^{\left[ 2 \right]}_{\mu\nu}$ in a more suitable form for different factorization. For instance, we can analyse the coefficient associated to the tensor  $\left( 4 u_{\mu}u_{\nu} +\eta_{\mu\nu} \right)$. For the perfect fluid, this coefficient is related to temperature $T$
\begin{align}
	T_{\mu\nu} \propto T^{d} \left( 4 u_{\mu}u_{\nu} +\eta_{\mu\nu} \right)
	\ .
	\label{BYmho00}
\end{align}
Setting $y=-\left( d / 4 \right)^{2}$ we have
\begin{align}
	-\frac{\mu}{2}\left[ 1 + \frac{d}{2} \lambda\left(\bB\cdot x\right) + \frac{1}{2} \left( \frac{d}{2}  \right)^{2} \lambda^{2} \left(\bB\cdot x\right)^{2} \right]
	\propto
	T^{d} \, \textrm{exp}\left[\frac{d}{2}\lambda \left( \bB \cdot x \right)\right]
	\ ,
	\label{BYmnho1}
\end{align}
where we reconstructed the series in the bilinears $\bB$. Doing this, the temperature of the fluid is modified as follows
\begin{align}
	T \longrightarrow T  \textrm{exp}\left[\frac{1}{2}\lambda \left( \bB \cdot x \right)\right]
	\ .
	\label{BYmho3}
\end{align}

\subsection{Stress--Energy Tensor for $AdS_4$}

The computation for the $AdS_{4}$ case is similar to the previous one. We consider the perturbative expansion
\begin{align}
	T_{\mu\nu} = -\frac{\mu}{2}\left( 3 u_{\mu}u_{\nu} +\eta_{\mu\nu} \right) + \lambda\mu\mathcal{T}^{\left[ 1 \right]}_{\mu\nu} + \lambda^{2}\mu\mathcal{T}^{\left[ 2 \right]}_{\mu\nu}
	\ ,
	\label{eta1LargeRTmunu4dim}
\end{align}
where we have defined $\mathcal{T}^{\left[ 1 \right]}$ as before and %check 2 in mho coefficient
\begin{eqnarray}
	\mathcal{T}^{\left[ 2 \right]}_{\mu\nu} &=& P^{\bot}_{\mu\nu}\left[\left(\frac{9}{8}+2y\right)\left(\bB\cdot x\right)^2 + \frac{21}{4}\left(x\bB\right)^{\bot}\left(x\bB\right)^{\shortparallel}\right] + \nonumber \\
&-&\frac{3}{8}P^{\shortparallel}_{\mu\nu}\left[\left(\frac{1}{2}+\frac{8}{3}y\right)\left(\bB\cdot x\right)^2+7\left(x\bB\right)^{\bot}\left(x\bB\right)^{\shortparallel}+2\left(\bB \bB\right)^{\bot}\left(xx\right)^{\shortparallel}\right]+ \nonumber \\
&+&y \left(\bB\cdot x\right)^2\left(\eta_{\mu\nu}+3u_{\mu}u_{\nu}\right) +
\frac{3}{4}\left[\left(\bB \bB\right)^{\bot}x_{\left(\mu\right.}x_{\left.\nu\right)}-2\left(x\bB\right)^{\bot}\bB_{\left(\mu\right.}x_{\left.\nu\right)}\right] + \nonumber \\
&-&\frac{3}{4} P^{\shortparallel}_{\left(\mu\rho\right.}P^{\bot}_{\left.\nu\right)\sigma}\left\{-3\bB^{\left[\rho\right.}x^{\left.\sigma\right]}\left[2 \bB \cdot x +\left(x\bB\right)^{\bot} \right]
-2x^{\rho}x^{\sigma}\left(\bB \bB\right)^{\bot}+\bB^{\rho}\bB^{\sigma}\left(xx\right)^{\shortparallel}\right\} \ , \nonumber \\
\end{eqnarray}

\section*{Conclusions}
%\addcontentsline{toc}{section}{Conclusions}

In present work, we have constructed the complete supersymmetric extension of classical solutions
of the $AdS$-Schwarzschild type. We denote the complete solution as the fermionic wig, to remind the
reader that the anticommuting nature of this ``hair''. We have provided the exact analytical solutions computed by
automatic elaboration and for that we have described the algorithm based on iterative solution of supergravity equations.
We are left out several interesting applications that will be studied in the future such as BPS solutions and their fermionic
wigs and in particular the applications to fluid/gravity correspondence.

\section*{Acknowledgements}

We are grateful to G. Policastro, L. Castellani and G. Dall'Agata for discussions during several stages of the present work.
We also would like to thank S. Minwalla, A. Marrani, L. Sommovigo and V. Pozzoli for very useful comments.

\appendix

\section{Definitions}

In this section we set the notations.

Characters $\left\{ A = 1,\cdots ,d \right\}$ labels bulk flat directions, indices $\left\{ M = 1,\cdots, d \right\}$ are bulk curved ones. Lowercase latin letters $\left\{ a\dots \right\}$ indicates flat boundary indices while greek lowercases $\left\{ \mu \dots\right\}$ are associated to curved boundary directions.

The flat metric is $\eta=\left\{ -,+,+,\dots \right\}$.

Products of gamma matrices are defined as follows
\begin{equation}
	\Gamma_{A_{1}\cdots A_{n}}= \Gamma_{\left[ A_{1} \right.}\cdots\Gamma_{\left. A_{n} \right]}
	\ ,
	\label{AlgGammaa}
\end{equation}
and antisymmetrizations read
\begin{equation}
	B_{\left[ A_{1}\cdots A_{n} \right]} = \frac{1}{n!}\left(
	B_{ A_{1}\cdots A_{n}}
	+
	\textrm{antisymmetrizations}
	\right) \ .
	\label{AlgAntisymm}
\end{equation}

%\section{Rarita-Schwinger Equation}
%
%In the paper \cite{Ceresole:2000jd}, the full action of $N=2$, $D=5$ gauged supergravity is given. Considering only the graviton and the massless gravitinos, the lagrangian $5$--form reduces to
%\begin{align}
%	\LC = &	
%	\frac{1}{2} R^{ab} \hat{e}_{ab}
%	+\frac{i}{4}e^{a}e^{b} \bar\psi^{i}\gamma_{ab} D\psi_{i}
%	-\frac{i}{32}e^{a}\bar\psi^{i}\gamma_{a}\psi_{i}\bar\psi^{j}\psi_{j}
%	\ ,
%	\label{DallAgataCeresole1}
%\end{align}
%where
%\begin{equation}
%	\hat{e}^{a_{1}\dots a_{p}} \equiv -\frac{1}{\left( 5-p \right)!}\varepsilon^{a_{1}\dots a_{p}b_{1}\dots b_{5}}e_{b_{1}}\dots e_{b_{5-p}}
%	\ .
%	\label{DallAgataCeresole2}
%\end{equation}
%The equation of motion for $\bar\psi^{i}_{\rho} $ reads
%\begin{align}
%	\varepsilon^{\mu\nu\rho\sigma\lambda}&
%	\left(
%	e^{a}_{\mu}e^{b}_{\nu}\gamma_{ab} D_{\sigma}\psi_{i\lambda}
%	+
%	\frac{1}{8}e^{a}_{\mu} \gamma_{a} \psi_{i\nu}\bar\psi^{j}_{\sigma}\psi_{j\lambda}
%	+
%	\frac{1}{8}e^{a}_{\mu} \bar\psi^{j}_{\nu}\gamma_{a} \psi_{j\sigma}\psi_{i\lambda}
%	\right) = 0
%	\ ,
%	\label{RS0}
%\end{align}
%which is the Rarita-Schwinger equation.

\section{Fierz Transformations}\label{FIERZ}

In this appendix we present the Fierz transformations used in this work.

The spinors used are $3$--dimensional Dirac spinors $\eta_{A}$ with $A$ labels the different spinors (and so, repeated $\left\{ A \right\}$ indices are not summed). It is easy to show that
\begin{align}
	\eta_{A} \eta_{A}{}^{\dagger}
	= &\
	-\frac{1}{2}
	\eta_{A}{}^{\dagger} \eta_{A} \hat\sigma_{0}
	-\frac{1}{2}
	\eta_{A}{}^{\dagger} \sigma_{i} \eta_{A} \hat\sigma^{i}
	\ .
	\label{Fierz0}
\end{align}
Using this, we get the following relations
\begin{align}
	\eta_{B}{}^{\dagger}\eta_{A}\eta_{A}{}^{\dagger}\eta_{B}
	=&\
	-\frac{1}{2} \eta_{B}{}^{\dagger}\eta_{B}\, \eta_{A}{}^{\dagger}\eta_{A}
	-
	\frac{1}{2} \eta_{B}{}^{\dagger}\hat\sigma_{i}\eta_{B}\, \eta_{A}{}^{\dagger}\hat\sigma^{i}\eta_{A}
	\ ,
	\label{Fierz1}
\end{align}
\begin{align}
	\eta_{B}{}^{\dagger}\eta_{A}\eta_{A}{}^{\dagger}\hat\sigma_{i}\eta_{B}
	=&\
	-
	\frac{1}{2} \eta_{A}{}^{\dagger}\eta_{A}\, \eta_{B}{}^{\dagger}\hat\sigma_{i}\eta_{B}
	-
	\frac{1}{2} \eta_{A}{}^{\dagger}\hat\sigma_{i}\eta_{A}\, \eta_{B}{}^{\dagger}\eta_{B}
	+	
	\frac{i}{2} \varepsilon_{ijk}\,  \eta_{B}{}^{\dagger}\hat\sigma^{j}\eta_{B}\, \eta_{A}{}^{\dagger}\hat\sigma^{k}\eta_{A}
	\ ,
	\label{Fierz2}
\end{align}
\begin{align}
	\eta_{B}{}^{\dagger}\hat\sigma_{i}\eta_{A}\eta_{A}{}^{\dagger}\hat\sigma_{j}\eta_{B}
	=&\
	-
	\frac{1}{2} \eta_{B}{}^{\dagger}\eta_{B}\, \eta_{A}{}^{\dagger}\eta_{A}\, \delta_{ij}
	+
	\frac{1}{2} \eta_{B}{}^{\dagger}\hat\sigma_{k}\eta_{B}\, \eta_{A}{}^{\dagger}\hat\sigma^{k}\eta_{A}\, \delta_{ij}
	+
	\nonumber\\ &
	-
	\frac{1}{2} \eta_{B}{}^{\dagger}\hat\sigma_{i}\eta_{B}\, \eta_{A}{}^{\dagger}\hat\sigma_{j}\eta_{A}
	-
	\frac{1}{2} \eta_{B}{}^{\dagger}\hat\sigma_{j}\eta_{B}\, \eta_{A}{}^{\dagger}\hat\sigma_{i}\eta_{A}
	+
	\nonumber\\ &
	-
	\frac{i}{2}\varepsilon_{ijk}\,  \eta_{A}{}^{\dagger}\eta_{A}\, \eta_{B}{}^{\dagger}\hat\sigma^{k}\eta_{B}
	+
	\frac{1}{2}\varepsilon_{ijk}\, \eta_{A}{}^{\dagger}\hat\sigma^{k}\eta_{A}\, \eta_{B}{}^{\dagger}\eta_{B}\,
	\ .
	\label{Fierz3}
\end{align}
If $A=B$ this reduces to
\begin{align}
	\eta{}^{\dagger}\hat\sigma_{i}\eta\eta{}^{\dagger}\hat\sigma_{j}\eta
	=&\
	-\frac{1}{2} \left( \eta{}^{\dagger}\eta \right)^{2}\,\delta_{ij}
	\ .
	\label{Fierz4}
\end{align}

%%%%%%%%%%%%%%%%%%%%%%%%%%%%%%%%%%%%%%%%%%%%%%%%%%%%%%%%%%%%%%%%%%%%%%%%%%%%%%%

\end{document}